\documentclass{jfm}
\usepackage{graphicx}
\usepackage{epstopdf, epsfig}
\usepackage{gensymb}
\usepackage{textcomp}
\usepackage{subfig}
\usepackage{amsmath}
\usepackage{hyperref}
\usepackage{cancel}
\usepackage{multirow}

\shorttitle{Turbulent scalar flux in inclined jets in crossflow}
\shortauthor{P. M. Milani, J. Ling and J. K. Eaton}

\title{Turbulent scalar flux in inclined jets in crossflow: counter gradient transport and deep learning modelling}

\author{Pedro M. Milani\aff{1}
  \corresp{\email{pmmilani@stanford.edu}},
  Julia Ling\aff{2}
 \and John K. Eaton\aff{1}}

\affiliation{\aff{1} Mechanical Engineering Department, Stanford University,
Stanford CA, USA
\aff{2} Citrine Informatics, Redwood City CA, USA}

\begin{document}

\maketitle

\begin{abstract}

A cylindrical and inclined jet in crossflow is studied under two distinct velocity ratios, $r=1$ and $r=2$, using highly resolved large eddy simulations (LES). First, an investigation of turbulent scalar mixing sheds light onto the previously observed but unexplained phenomenon of negative turbulent diffusivity. We identify two distinct types of counter gradient transport, prevalent in different regions: the first, throughout the windward shear layer, is caused by cross-gradient transport; the second, close to the wall right after injection, is caused by non-local effects. Then, we propose a deep learning approach for modelling the turbulent scalar flux by adapting the tensor basis neural network previously developed to model Reynolds stresses \citep{julia_deepnn}. This approach uses a deep neural network with embedded coordinate frame invariance to predict a tensorial turbulent diffusivity that is not explicitly available in the high fidelity data used for training. After ensuring analytically that the matrix diffusivity leads to a stable solution for the advection diffusion equation, we apply this approach in the inclined jets in crossflow under study. The results show significant improvement compared to a simple model, particularly where cross-gradient effects play an important role in turbulent mixing. The model proposed herein is not limited to jets in crossflow; it can be used in any turbulent flow where the Reynolds averaged transport of a scalar is considered.

\end{abstract}


\section{Introduction}

The jet in crossflow is a canonical geometry in fluid mechanics, whereby fluid is ejected from an orifice and interacts with main flow passing over that orifice \citep{mahesh_review2013}. This configuration has several practical applications. In the turbomachinery community, film cooling systems eject cooler air through rows of holes in the blade surface, which ideally stays close to the surface and protects it against the hot mainstream \citep{bogard_review}. In atmospheric flows, some types of volcanic plumes rise from the vent while interacting with prevailing winds. This interaction helps determine mixing and deposition of ash \citep[e.g.][]{gardner2007eruption}. Most interesting jets in crossflow involve mixing because the jet carries a scalar contaminant, such as heat or particles, into the crossflow. In these problems, it is imperative to analyze the resulting scalar concentration field, denoted by $c$, in addition to the velocity field $u_i$.

Different parameters determine the resulting flow in a jet in crossflow, including geometry (such as hole shape and angle), Reynolds number, and crossflow characteristics (like incoming boundary layer thickness and turbulence). One key dimensionless parameter in the incompressible regime is the velocity ratio $r$, defined as

\begin{equation}
\label{eq-r}
r=U_j/U_c,  
\end{equation}

\noindent where $U_j$ is the bulk velocity of the jet and $U_c$ is the bulk velocity of the crossflow. In general, low velocity ratio jets stay close to the wall, while high velocity ratio jets penetrate deep into the crossflow before turning.

The literature on jets in crossflow is extensive \citep{mahesh_review2013}. \citet{fric1994vortical} used flow visualization and qualitatively described four types of vortices that appear in transverse jets in crossflow: the counter-rotating vortex pair, jet-shear layer vortices, horsehoe vortices, and wake vortices. \citet{su2004simultaneous} and \citet{schreivogel2016simultaneous} used different techniques to simultaneously measure velocity and concentration, and then reported some turbulent mixing statistics. Computational studies have applied the Reynolds-averaged Navier-Stokes (RANS) equations, but scalar transport is not computed accurately due to deficiencies in the available turbulence models \citep{acharya2001flow}. High-fidelity simulations, on the other hand, have been shown to be much more predictive. \citet{yuan1999large} performed one of the first successful large eddy simulations (LES) of a jet in crossflow. \citet{muppidi2008direct} reported results from a direct numerical simulation (DNS) that agreed well with the experiment of \citet{su2004simultaneous}, a transverse jet in crossflow at $r=5.7$. They analyzed entrainment rates and reported on inadequacies of standard turbulent scalar mixing models, including regions of perceived counter gradient transport.

The turbulence community has dedicated much effort to the modelling of the Reynolds stress, and multiple models are available and commonly employed in jets in crossflow \citep{hoda1999predictions}. Previous research also shows that errors in the turbulent mixing models contribute decisively to poor mean scalar field predictions in these flows \citep{julia_analysisJICF}. Therefore, the present paper focuses solely on modelling the turbulent scalar flux, $\overline{u_i'c'}$, which represents the effect of turbulence on the mean scalar concentration field and is unclosed in the Reynolds-averaged scalar transport equation.

The most commonly used model, called the gradient diffusion hypothesis (GDH), assumes that turbulence acts like isotropic diffusion on the mean concentration gradient as shown in

\begin{equation}
\overline{u_i'c'} = -\alpha_t \frac{\partial \bar{c}}{\partial x_i}.
\label{eq-gdh}
\end{equation}

\noindent The spatially varying turbulent diffusivity field in eq.~\ref{eq-gdh} is $\alpha_t$, which is usually specified by means of a turbulent Schmidt number, $Sc_t \equiv \alpha_t / \nu_t$, where the eddy viscosity $\nu_t$ is given by a baseline momentum model \citep{kays1994turbulentPrt, combest2011gradient}.

More advanced models have been proposed, but have not gained much traction outside of research laboratories \citep{combest2011gradient}. \citet{batchelor1949diffusion} recognized that a tensor diffusivity $D_{ij}$ is more appropriate:

\begin{equation}
\overline{u_i'c'} = -D_{ij} \frac{\partial \bar{c}}{\partial x_j}.
\label{eq-gdhtensor}
\end{equation}

\noindent Different authors \citep[e.g.][]{tavoularis1981experiments} measured diagonal and off-diagonal elements of $D_{ij}$ in simple flows. To actually solve the scalar equation, \citet{daly1970transport} and \citet{abe2001towards} proposed algebraic models where $D_{ij}$ is specified based on the Reynolds stress. Such formulations can theoretically improve results, but recent work in jets in crossflow has shown that they are not necessarily superior to simpler models and struggle with convergence issues \citep{julia_analysisJICF, ryan2017skewed}.

Previous work highlighted some deficiencies of the GDH of eq.~\ref{eq-gdh} in jets in crossflow \citep[e.g.][]{weatheritt2020data, kohli2005turbulent}. In the present paper, we focus our attention to a particular observation that has been reported before: regions of negative turbulent diffusivity, or counter gradient transport. This means that at least one component of the $\overline{u_i'c'}$ vector has the same sign as the equivalent component of $\frac{\partial \bar{c}}{\partial x_i}$, which directly contradicts eq.~\ref{eq-gdh} with a positive turbulent diffusivity. This was observed in jets in crossflow by several authors, both numerically \citep{muppidi2008direct, salewski2008mixing, bodart2013highfidelity, milani2018magnetic} and experimentally \citep{salewski2008mixing, schreivogel2016simultaneous}. These workers reported the phenomenon without discussing its causes, so the present paper expands their analysis by explaining the observed counter gradient transport. This also inspired us to devise a deep learning approach for modelling the turbulent scalar flux that does not rely on the simple GDH of eq.~\ref{eq-gdh}.

Machine learning tools have been rising in popularity in the turbulence closure literature, as evidenced by the review of \citet{duraisamy2019turbulence}. \citet{julia_deepnn} used deep neural networks to predict coefficients in a basis expansion for the Reynolds stress anisotropy tensor, which became known as tensor basis neural network (TBNN). \citet{parish2016paradigm} proposed a modelling strategy that couples machine learning with field inversion through adjoint optimization to develop new models. \citet{sandberg2018applying} used a genetic optimization algorithm to generate data-driven analytical expressions for turbulent anisotropy and turbulent diffusivity and \citet{maulik2019sub} explored the use of a neural network towards subgrid scale modelling in large eddy simulations. The bulk of the aforementioned literature focuses on Reynolds stress modeling, but some authors have had success in applying machine learning techniques for closure of the passive scalar equation \citep{milani2018approach, sotgiu2018turbulent, sandberg2018applying, weatheritt2020data, milani2020generalization}. In the present paper, a different scalar closure model is proposed, leveraging the tensor basis neural network concept of \citet{julia_deepnn} to generate a more general model form.

The contribution of the present paper is twofold. First, we analyze and classify the regions of counter gradient transport found in an inclined jet in crossflow. These results suggest that an anisotropic model is needed to better capture the turbulent transport in part of the flow. Second, we adapt the popular TBNN methodology, which originally predicts the Reynolds stress tensor, to model the turbulent scalar flux vector. We analyze the results of this approach, denoted TBNN-s (where the ``s" stands for scalar flux modelling), with the counter gradient transport regions in mind. The paper is organized as follows: section 2 describes the flow under consideration and the high fidelity simulations used to generate the data; section 3 identifies and analyzes the counter gradient transport regions; section 4 contains the description and the results of the deep learning model employed; and section 5 presents conclusions and suggestions for future work.

\section{Numerical setup}

Two highly resolved large eddy simulations (LES) are used in the present work. The simulations were described in detail and validated against experimental data in \citet{milani2019enriching}. The present paper employs those results for analysis, thus this section provides only a brief summary.

\begin{figure}
  \begin{center}
  \includegraphics[width = 120mm]{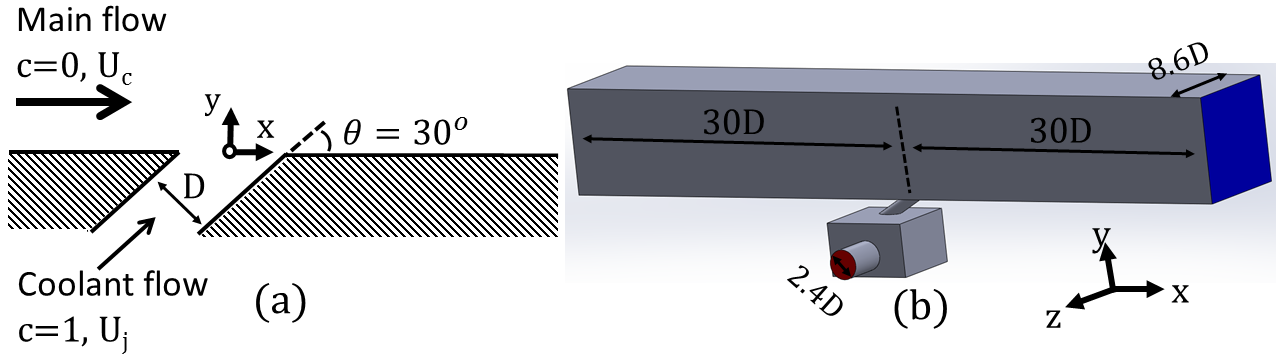}
  \end{center}
\caption{(a) Schematic of the inclined jet in crossflow. The coordinate frame origin is located at the center of the hole as it meets the bottom wall. (b) 3D view of the complete simulation domain, with plenum feed inlet plane shown in red, outlet plane in blue, and walls in grey. Figures adapted from \citet{milani2019enriching}.}
\label{fig-1-schematicsdomain}
\end{figure}

The geometry considered is a fully 3D inclined jet in crossflow, shown in figure~\ref{fig-1-schematicsdomain}(a). The hole has a circular cross-section of diameter $D$ and is inclined $30 ^\circ$ with respect to the streamwise direction. The mainstream consists of a turbulent developing flow in a square channel, whose boundary layer thickness just upstream of the hole location is $\delta_{99}/D = 1.5$. The freestream turbulence level is roughly $1 \%$. The total simulation domain is shown in figure~\ref{fig-1-schematicsdomain}(b). Note that the injection hole is short, with length $4.1D$, and is fed from a rectangular plenum underneath the channel. This configuration was chosen partly because it is representative of film cooling applications; it also means that the flow inside the hole is highly non-uniform and contains secondary flows \citep{milani2019enriching}.

The incompressible, filtered continuity and Navier-Stokes equations are solved in 3D as shown in eq.~\ref{eq-ns}. The tilde over a variable is shown explicitly to denote filtered quantities; $u_i$ are the Cartesian components of the velocity, and $p$ is the pressure. The unresolved scales are accounted for in the subgrid scale stress, $\tilde{\tau}_{ij} = \widetilde{u_i u_j} - \tilde{u}_i \tilde{u}_j$, which is closed using the Vreman model \citep{vreman2004eddy}. The fluid density $\rho$ and kinematic viscosity $\nu$ are constant. The two simulations use the same geometry and main channel inlet condition, and their difference lies in the velocity ratio $r$, which is varied by changing the flow rate into the plenum. The resulting jet Reynolds number is $Re_D = U_j D / \nu = 2,900$ for the $r=1$ case and $Re_D=5,800$ for $r=2$.

\begin{equation}
\label{eq-ns}
\frac{\partial \tilde{u}_k}{\partial x_k} = 0; \qquad \frac{\partial \tilde{u}_i}{\partial t} + \frac{\partial \left( \tilde{u}_j \tilde{u}_i \right)}{\partial x_j} = -\frac{1}{\rho} \frac{\partial \tilde{p}}{\partial x_i} + \nu \frac{\partial^2 \tilde{u}_i}{\partial x_j \partial x_j} - \frac{\partial}{\partial x_j} \tilde{\tau}_{ij}
\end{equation}

The filtered equation for a passive scalar $c$ is also solved as shown in eq.~\ref{eq-scalar}. The molecular Schmidt number is $Sc=1$ and the subgrid scale mixing is represented by $\tilde{\sigma}_{j} = \widetilde{c u_j} - \tilde{c} \tilde{u}_j$, which is modelled employing the Reynolds analogy with a constant turbulent Schmidt number, $Sc_t=0.85$. The scalar is injected with concentration $c=1$ in the jet and $c=0$ in the mainstream; all walls are adiabatic.

\begin{equation}
\label{eq-scalar}
\frac{\partial \tilde{c}}{\partial t} + \frac{\partial \left( \tilde{u}_j \tilde{c} \right)}{\partial x_j} = \frac{\nu}{Sc} \frac{\partial^2 \tilde{c}}{\partial x_j \partial x_j} - \frac{\partial}{\partial x_j} \tilde{\sigma}_{j}
\end{equation}

The equations are solved with the incompressible solver Vida from Cascade Technologies. It employs the method developed by \citet{ham2007efficient} which consists of second-order accurate spatial discretization and explicit time advancement, based on second-order Adams Bashforth. The mesh is block-structured and contains exclusively hexahedral elements, either $40.1M$ ($r=1$) or $48.3M$ ($r=2$). The $y^+$ values in the hole and bottom walls are around $1$, so no wall models are employed there. Throughout the whole jet-crossflow interaction, the subgrid scale viscosity is significantly smaller than the molecular viscosity, so the present LES's have almost DNS-like resolution in the regions of interest. Therefore we subsequently omit the tilde over grid-filtered quantities. Thus, $u_i$ and $c$ will refer to the (filtered) results from the LES, overbars will represent time averages of filtered quantities, and primes will denote fluctuating quantities in the filtered sense (so $c' \equiv c - \bar{c}$). The inlet conditions are generated using a method based on that from \citet{xie2008efficient} and are chosen to match a concomitant experimental setup. The validation is performed against 3D magnetic resonance imaging (MRI) data for both velocity and scalar concentration and the agreement is excellent. More details including grid and time convergence studies are in \citet{milani2019enriching}.

\begin{figure}
  \begin{center}
  \includegraphics[width = 130mm]{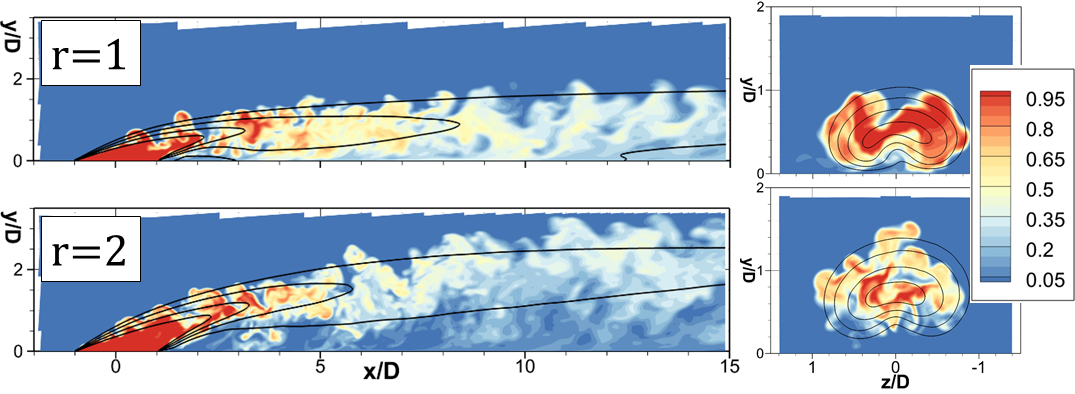}
  \end{center}
\caption{Scalar concentration results $c$ on the centerplane ($z=0$, left) and in an axial plane located at $x/D=2$ (right). Color contours show an instantaneous snapshot of $c$ and lines indicate isocontours of mean concentration, $\bar{c}=0.2, 0.4, 0.6, 0.8$.}
\label{fig-2-csummary}
\end{figure}

Figure~\ref{fig-2-csummary} shows contour plots of scalar concentration $c$ in the $r=1$ and $r=2$ cases to provide some physical insight into the flows. In both cases, the jet detaches from the bottom wall as soon as it is injected; it re-attaches in the lower velocity ratio case, but remains detached under the higher velocity ratio. As the $y-z$ planes show, the mean profile assumes a typical kidney shape due to the influence of the counter rotating vortex pair (CVP), a distinguishing feature of this flow \citep{mahesh_review2013}. More detailed results are in \citet{milani2019enriching}.

\section{Counter gradient transport}

In this section, we use the validated high-fidelity simulations to study turbulent mixing. The focus is two distinct regions of negative turbulent diffusivity (or counter gradient transport) which we classify as \textit{Type 1} and \textit{Type 2}.

\subsection{Type 1}

\begin{figure}
  \begin{center}
    \subfloat[$r=1$]{{\includegraphics[width=42mm]{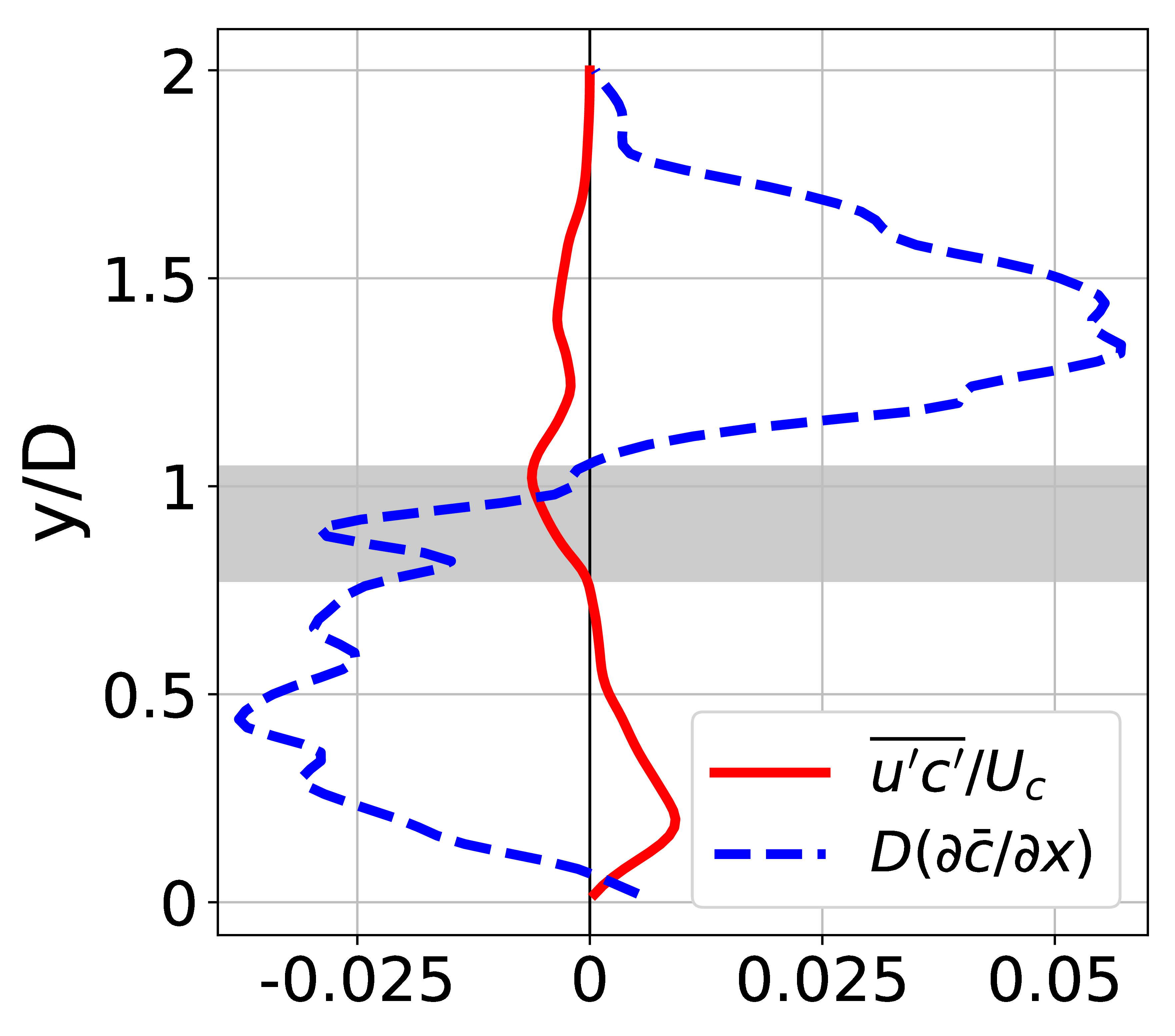} }}%
    \subfloat[$r=1$]{{\includegraphics[width=42mm]{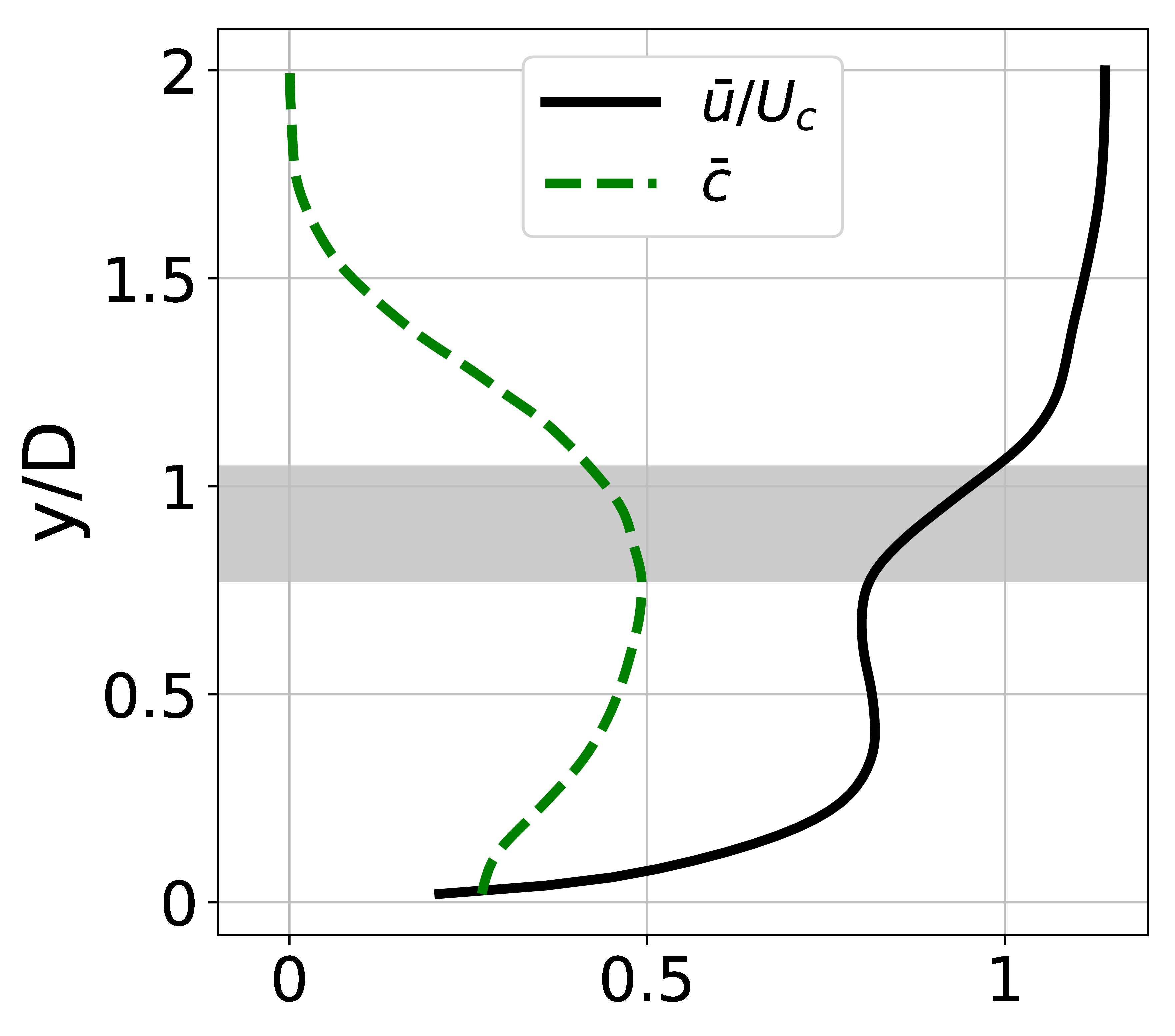} }}\\
    \subfloat[$r=2$]{{\includegraphics[width=42mm]{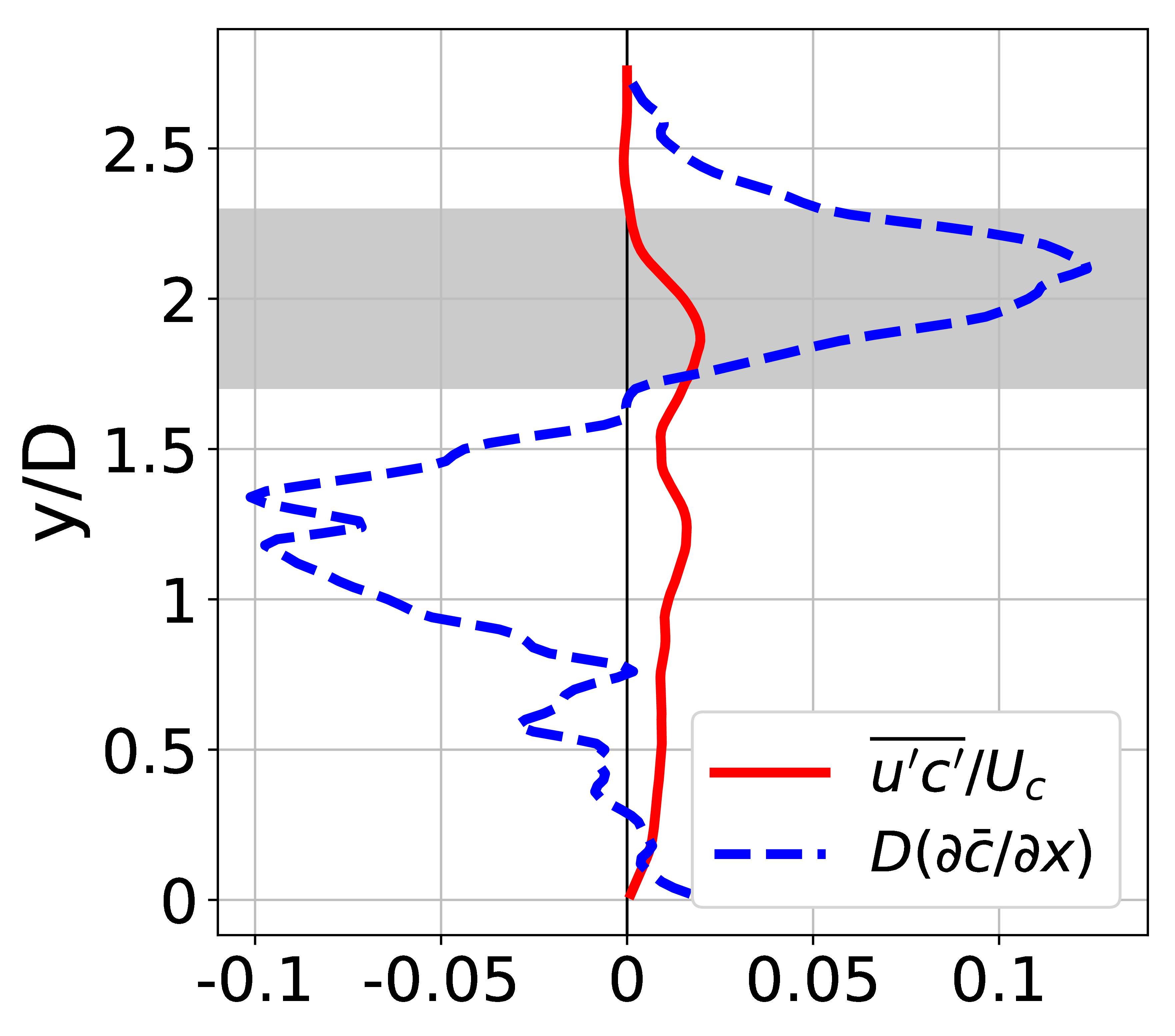} }}%
    \subfloat[$r=2$]{{\includegraphics[width=42mm]{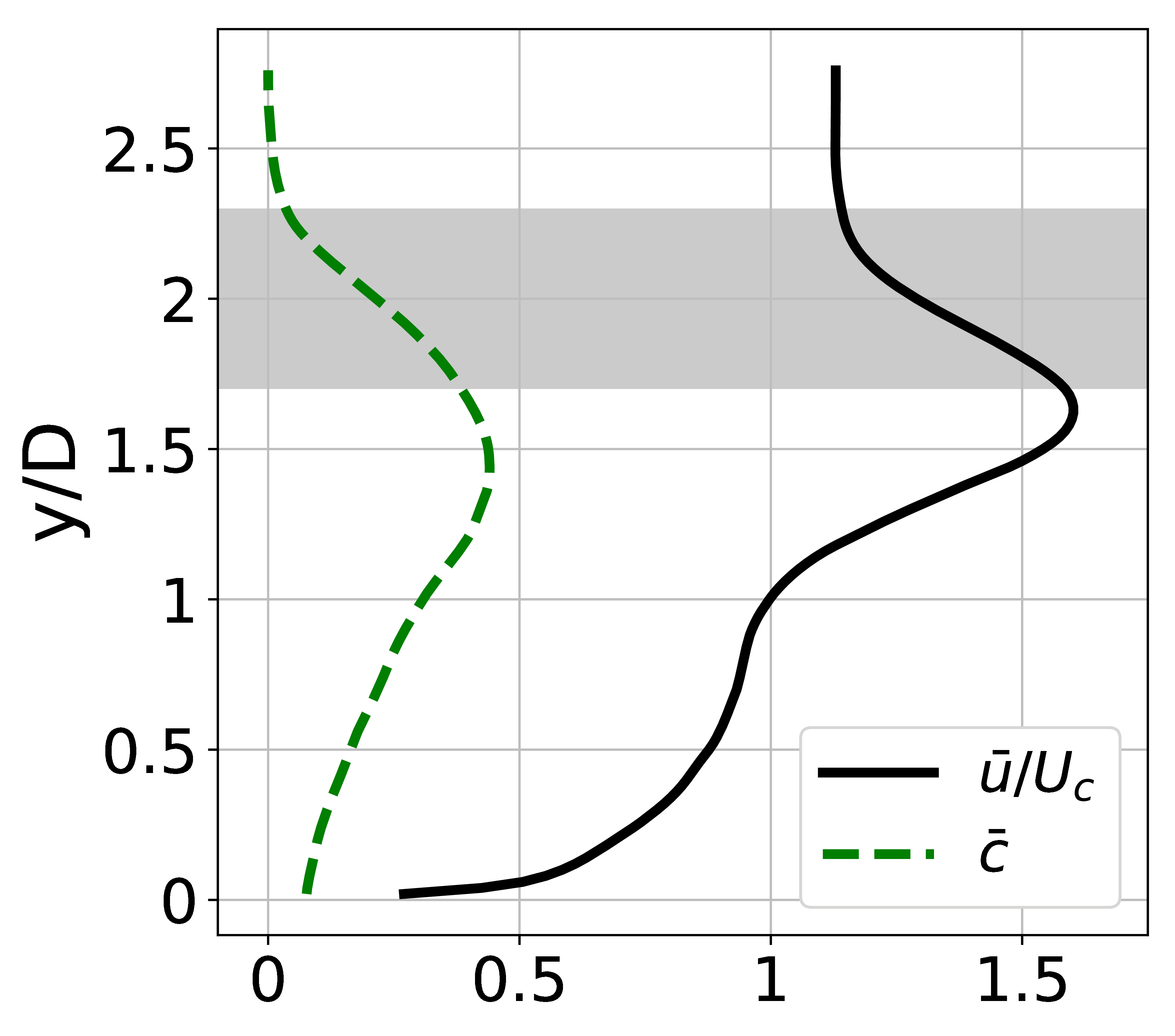} }}%
  \end{center}
\caption{Vertical profiles at $x/D=5$ and $z/D=0$ from the $r=1$ (top row) and $r=2$ (bottom row) LES. (a) and (c) show streamwise turbulent scalar flux and mean concentration gradient. (b) and (d) show mean streamwise velocity and concentration. The grey area shows a region of implied negative diffusivity in the streamwise direction.}
\label{fig-3-negativediff_uc}
\end{figure}

Figure~\ref{fig-3-negativediff_uc}(a) shows vertical profiles of the streamwise components, $\overline{u'c'}$ and $\partial \bar{c} / \partial x$ for $r=1$. These are located at $x/D=5$ on the centerline ($z/D=0$). There is a region, marked in grey, where both $\overline{u'c'} < 0$ and $\partial \bar{c} / \partial x < 0$. Figure~\ref{fig-3-negativediff_uc}(b) presents profiles of mean streamwise velocity and concentration, $\bar{u}$ and $\bar{c}$, in the same location. It shows that the region with implied streamwise counter gradient transport is located just above the jet centerline, which is a similar observation to that of \citet{schreivogel2016simultaneous} and \citet{bodart2013highfidelity}. Here, the likely cause of that phenomenon is not turbulent motion in the $x$ direction, but instead turbulent motion in the $y$ direction. In the grey area, turbulent eddies of sufficiently small size that bring fluid from above ($v' < 0$) tend to bring weaker concentration ($c' < 0$) and higher streamwise velocity ($u' > 0$) since locally $\partial \bar{c} / \partial y < 0$ and $\partial \bar{u} / \partial y > 0$. Similarly, eddies that bring fluid from below ($v' > 0$) correlate $c' > 0$ with $u' < 0$. That causes $\overline{u'c'}$ to be negative, which implies negative diffusivity since $\partial \bar{c} / \partial x < 0$.

The same phenomenon manifests itself in higher velocity ratio jets, with slight differences. Figure~\ref{fig-3-negativediff_uc}(c) shows vertical profiles at $x/D=5$ and $z/D=0$ from the $r=2$ LES. As is clear in figure~\ref{fig-3-negativediff_uc}(d), the jet core is faster than the free stream flow, so the $y$ gradients of mean streamwise velocity and mean scalar gradient are roughly aligned in the top half of the jet, unlike when $r=1$. Through a similar argument as before, vertical turbulent transport produces positive $\overline{u'c'}$ in that region. Thus, the perceived negative diffusivity region should occur when $\partial \bar{c} / \partial x > 0$. The streamwise mean scalar gradient is controlled by two main, competing effects: the spreading of the jet (which causes the concentration in the core to decrease), and the fact that the jet as a whole is moving up, away from the wall. Positive mean concentration streamwise gradients do not happen immediately above the jet centerline, but instead are present further up in the windward shear layer, as seen in figure~\ref{fig-3-negativediff_uc}(c). So, the grey band is located higher up and is wider when compared to the one shown in figure~\ref{fig-3-negativediff_uc}(a) for $r=1$.

\begin{figure}
  \begin{center}
    \subfloat[$r=1$]{{\includegraphics[width=45mm]{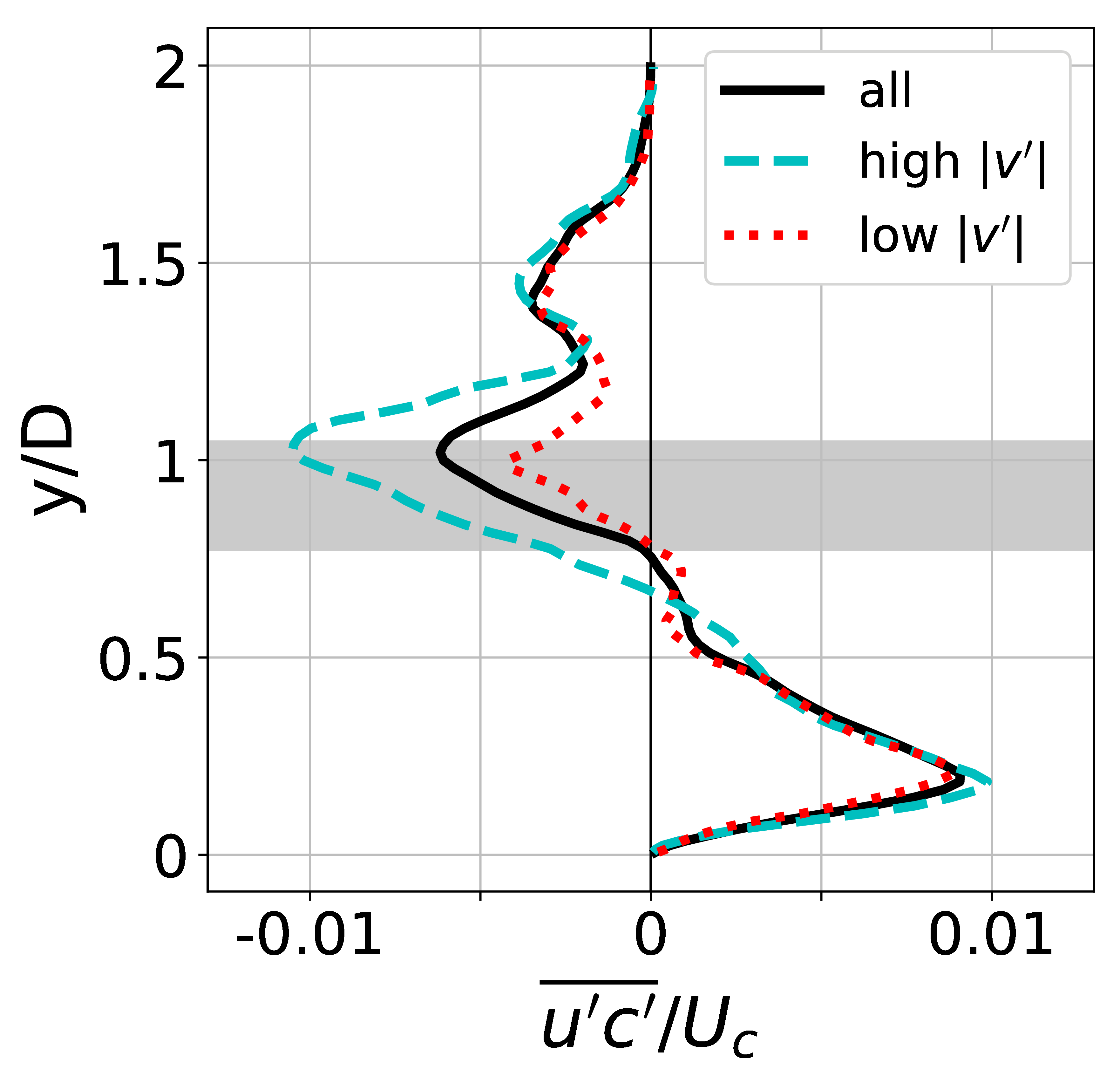} }}%
    \subfloat[$r=2$]{{\includegraphics[width=45mm]{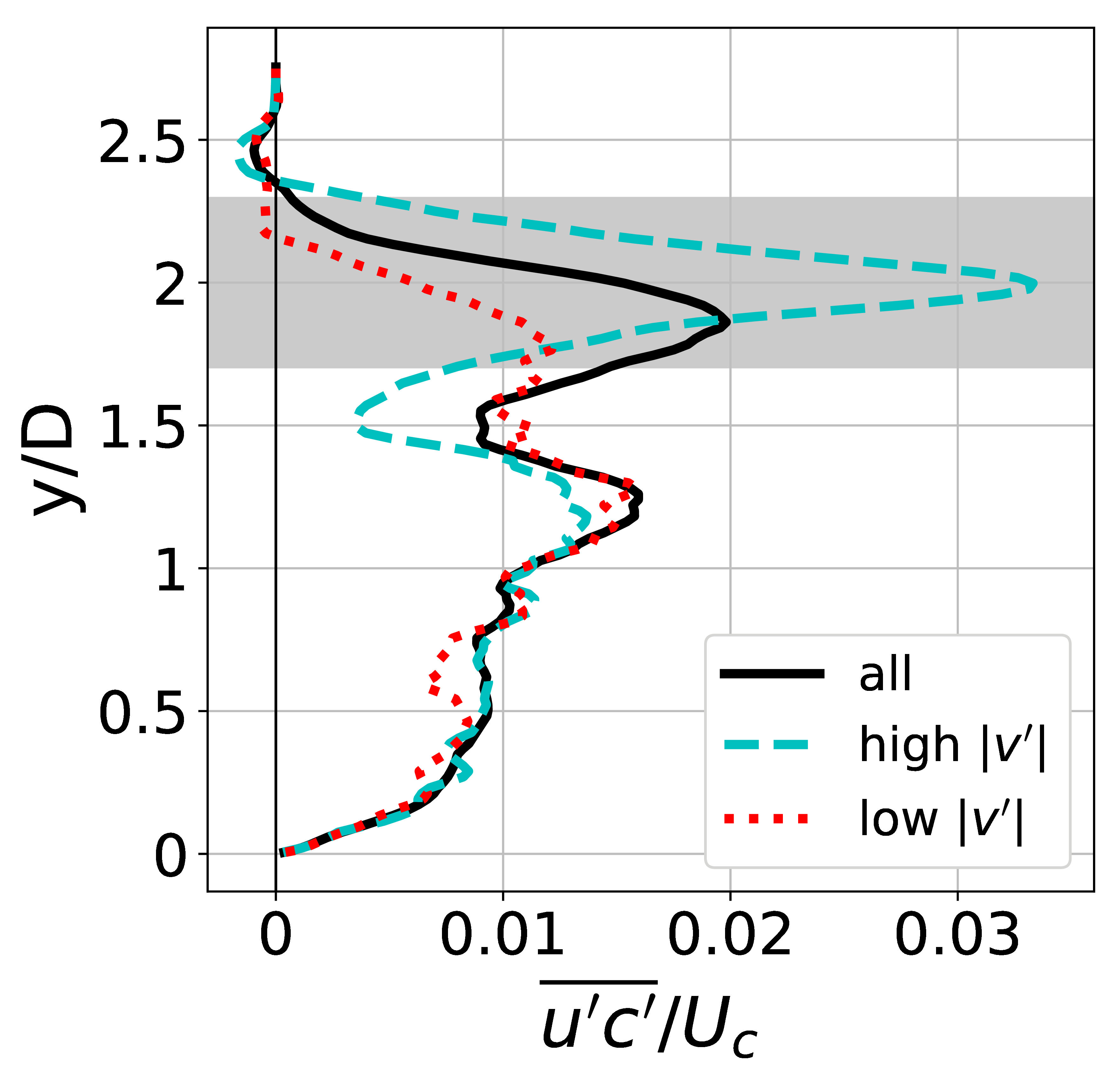} }}%
  \end{center}
\caption{Vertical profiles at $x/D=5$ and $z/D=0$ of $\overline{u'c'}$ from the LES. Different averages are performed: unconditional, or conditioned on high/low values of $|v'|$ at the center of the grey band.}
\label{fig-4-negativediff_cond}
\end{figure}

To support these explanations, consider figures~\ref{fig-4-negativediff_cond}(a)-(b). They show three differently averaged profiles of $\overline{u'c'}$ in the same location as shown in figure~\ref{fig-3-negativediff_uc}, for $r=1$ and $r=2$. For the solid line, the quantity $u'c'$ is averaged over all available time steps to produce the same quantity shown in figure~\ref{fig-3-negativediff_uc}(a) and \ref{fig-3-negativediff_uc}(c). For the other two lines, the average is conditioned on the local value of the magnitude of the vertical velocity fluctuation, $|v'|$, at the center of the grey band ($y/D=0.91$ and $y/D=2$ respectively). The average is taken either when $|v'|$ is high (top 20\% of samples) or low (bottom 20\% of samples). Away from the shaded region, the three lines converge as expected since the value of $v'$ is uncorrelated with $u'c'$ at long distances. However, in the grey band, the three lines are significantly different: $\overline{u'c'}$ has much higher magnitude when $|v'|$ is high, and much smaller magnitude when $|v'|$ is low. This shows that vertical transport indeed plays an important role in setting the value of $\overline{u'c'}$ in this region and that events of high $|v'|$ accentuate the perceived negative diffusivity in the streamwise direction.

To investigate the behavior of $\overline{u'_i c'}$ in more detail, it is useful to examine its transport equation, shown in eq.~\ref{eq-uctransport}. This equation is exact, and can be derived by multiplying $u'_i$ to the (unaveraged) scalar transport equation and $c'$ to the (unaveraged) Navier-Stokes equation, adding the results, and performing a time average. Capital letters $U$ and $C$ stand for time averages exactly like overbars (thus $\bar{c} = C$), but are used here for a leaner notation. $\delta_{ij}$ is the Kronecker delta and $\alpha$ is the molecular diffusivity of the scalar, so $\alpha = \nu / Sc$. Note that these equations are mostly analogous to the Reynolds stress transport equations. Term A is the advection due to the mean velocity. Term I is the production term and shows how mean velocity and scalar gradients can locally generate $\overline{u'_i c'}$. Term II is the viscous destruction term. Term III consists of different mechanisms of turbulent transport, and term IV is a source term due to the pressure-scalar gradient correlation. For more details and modelling ideas, consult \citet{younis2005rational}.

\begin{multline}
\label{eq-uctransport}
\frac{\partial \overline{u'_i c'}}{\partial t} + \overbrace{ \frac{ \partial \left( U_j \overline{u'_i c'}\right) }{\partial x_j}}^\text{A} = \overbrace{-\overline{u'_j c'} \frac{\partial U_i}{\partial x_j} - \overline{u'_j u'_i} \frac{\partial C}{\partial x_j}}^\text{I} \overbrace{ - (\nu + \alpha)\overline{\frac{\partial c'}{\partial x_j} \frac{\partial u'_i}{\partial x_j}}}^\text{II} \\
\overbrace{-\frac{\partial}{\partial x_j} \left( \overline{u'_j u'_i c'} + \frac{\overline{p'c'}}{\rho} \delta_{ij} - \alpha \overline{u'_i \frac{\partial c'}{\partial x_j}} - \nu \overline{c' \frac{\partial u'_i}{\partial x_j}} \right)}^\text{III} \overbrace{+ \overline{\frac{p'}{\rho} \frac{\partial c'}{\partial x_i}}}^\text{IV}
\end{multline}

\begin{figure}
  \begin{center}
    \subfloat[$r=1$]{{\includegraphics[width=52.5mm]{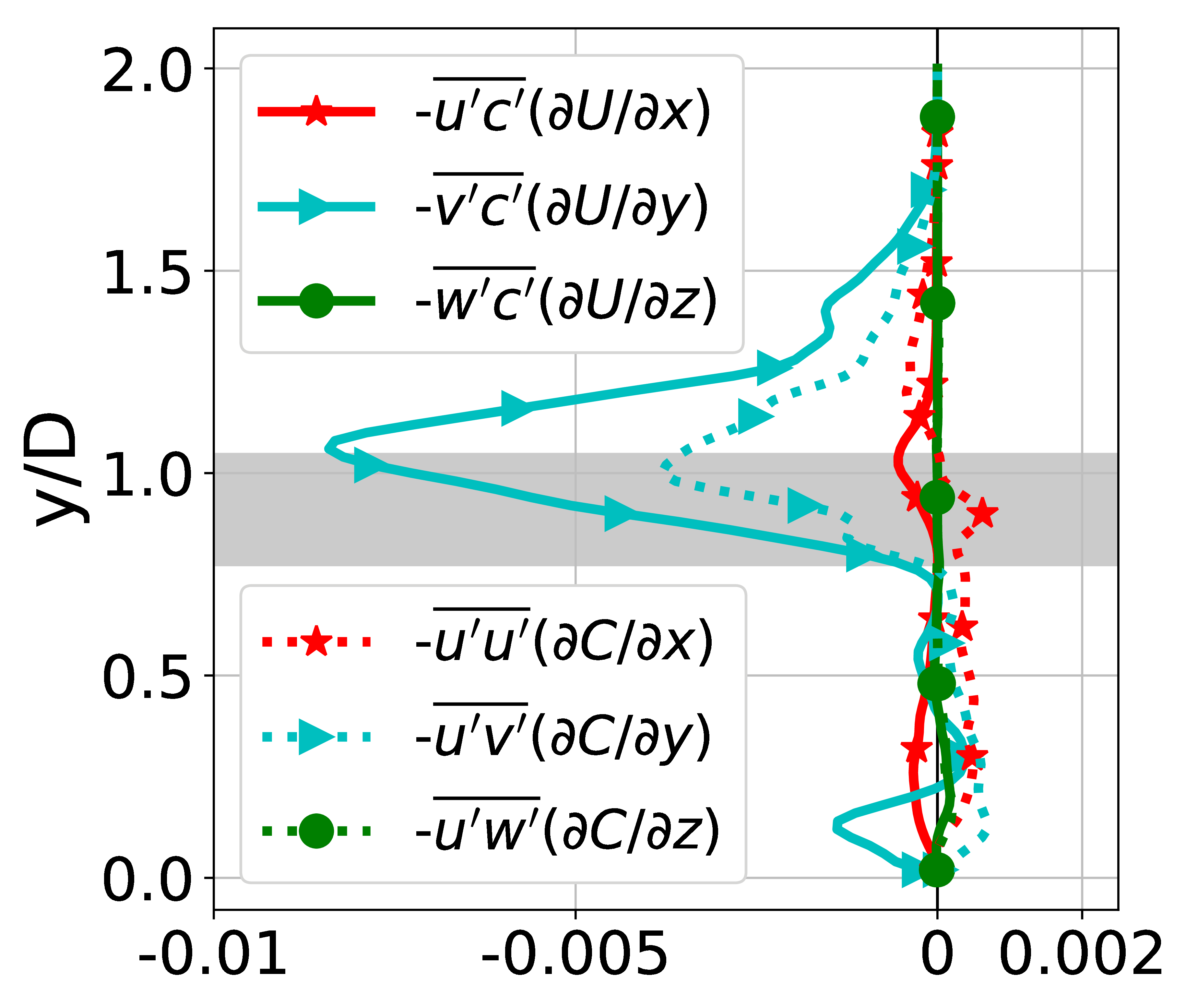} }}%
    \subfloat[$r=2$]{{\includegraphics[width=52.5mm]{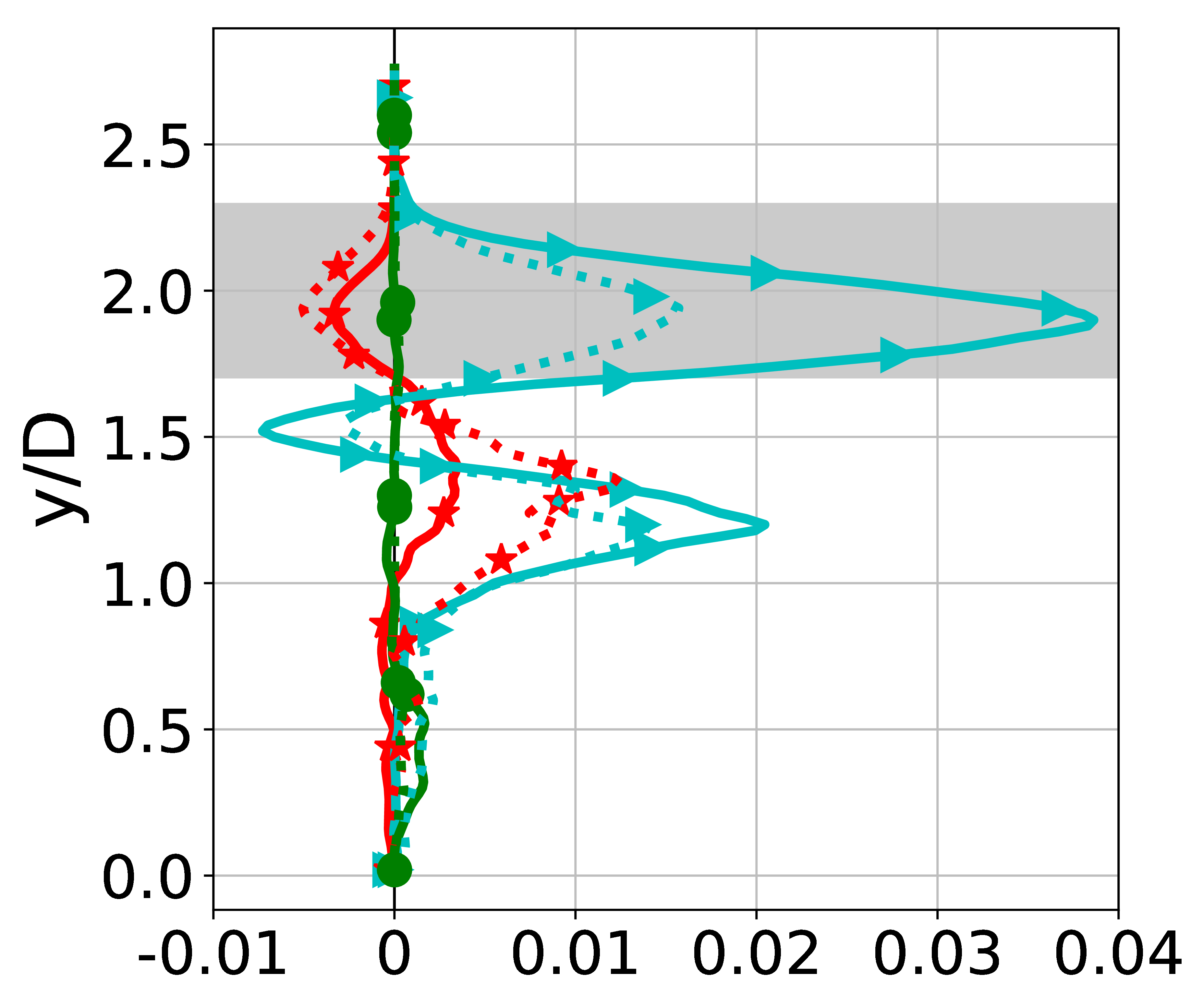} }}%
  \end{center}
\caption{Vertical profiles of different components of the $\overline{u'c'}$ production term (marked as I in eq.~\ref{eq-uctransport}). The profiles are shown at $x/D=5$, $z/D=0$ and the grey band represents the area of negative diffusivity. Terms are non-dimensionalized by $U_c$ and $D$.}
\label{fig-5-production_uc}
\end{figure}

As further evidence that vertical velocity and concentration gradients are responsible for the streamwise negative diffusivity shown in figure~\ref{fig-3-negativediff_uc}, we consider the production terms of eq.~\ref{eq-uctransport} for $\overline{u'c'}$. In figure~\ref{fig-5-production_uc}, vertical profiles of the six different components of term I are shown. Note that the gradient diffusion hypothesis (GDH) considers only the effects of the mean concentration gradient in the x-direction on $\overline{u'c'}$, which is reflected on the production term $-\overline{u'u'} \frac{\partial C}{\partial x}$. However, in the regions where counter gradient transport is observed, the production term is dominated by the components containing the vertical gradients: $-\overline{v'c'} \frac{\partial U}{\partial y}$ and $-\overline{u'v'} \frac{\partial C}{\partial y}$. These terms are indeed negative in figure~\ref{fig-5-production_uc}(a) and positive in figure~\ref{fig-5-production_uc}(b), which explains the sign of $\overline{u'c'}$ in the grey regions.

There is one important caveat in our discussion of Type 1 counter gradient transport. In shear flows, the resulting mean scalar concentration is not a strong function of the modelled $\overline{u'c'}$ since the mean advection typically dominates the scalar transport in the streamwise direction. This point is discussed in more detail in appendix~A. So, a turbulence model that fails to capture this phenomenon might still produce acceptable mean scalar field results in jets in crossflow. We believe the present analysis is still relevant for two reasons. First, it builds physical understanding of the turbulent transport in 3D jets in crossflow by showing a concrete instance of the failure of the GDH and explaining the underlying cause. Such understanding is necessary to design improved and robust mixing models, and could be useful to interpret turbulence results in other flows as well. Second, in different contexts the correct prediction of these Type 1 transport regions might be more relevant. For example, cross-gradient transport could be more important in different 3D turbulent flows; or the quantity of interest might not be the mean scalar field, and instead be more sensitive to the streamwise scalar flux.

\subsection{Type 2}

\begin{figure}
  \begin{center}
    \subfloat[$r=1$]{{\includegraphics[width=65mm]{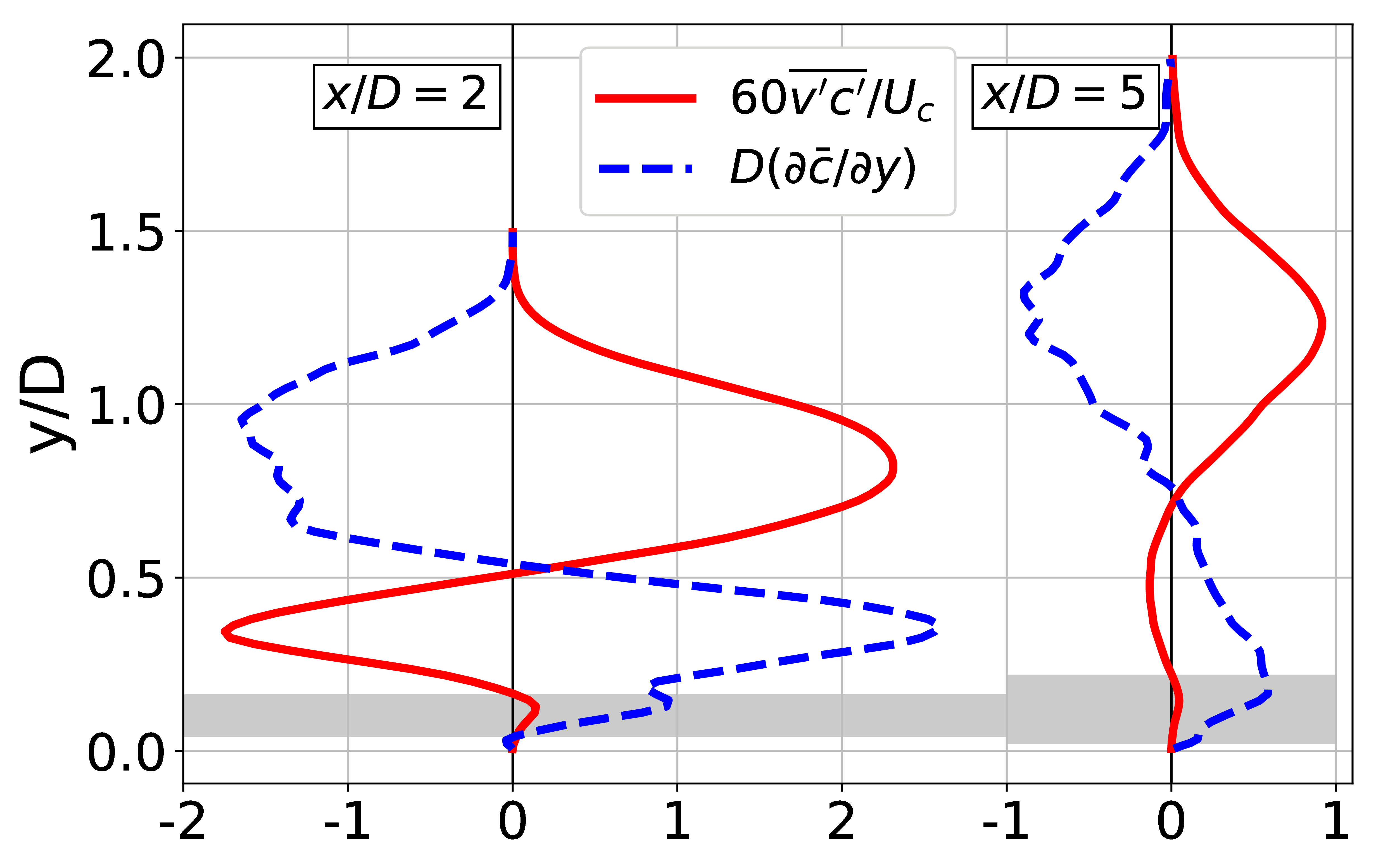} }}%
    \subfloat[$r=2$]{{\includegraphics[width=65mm]{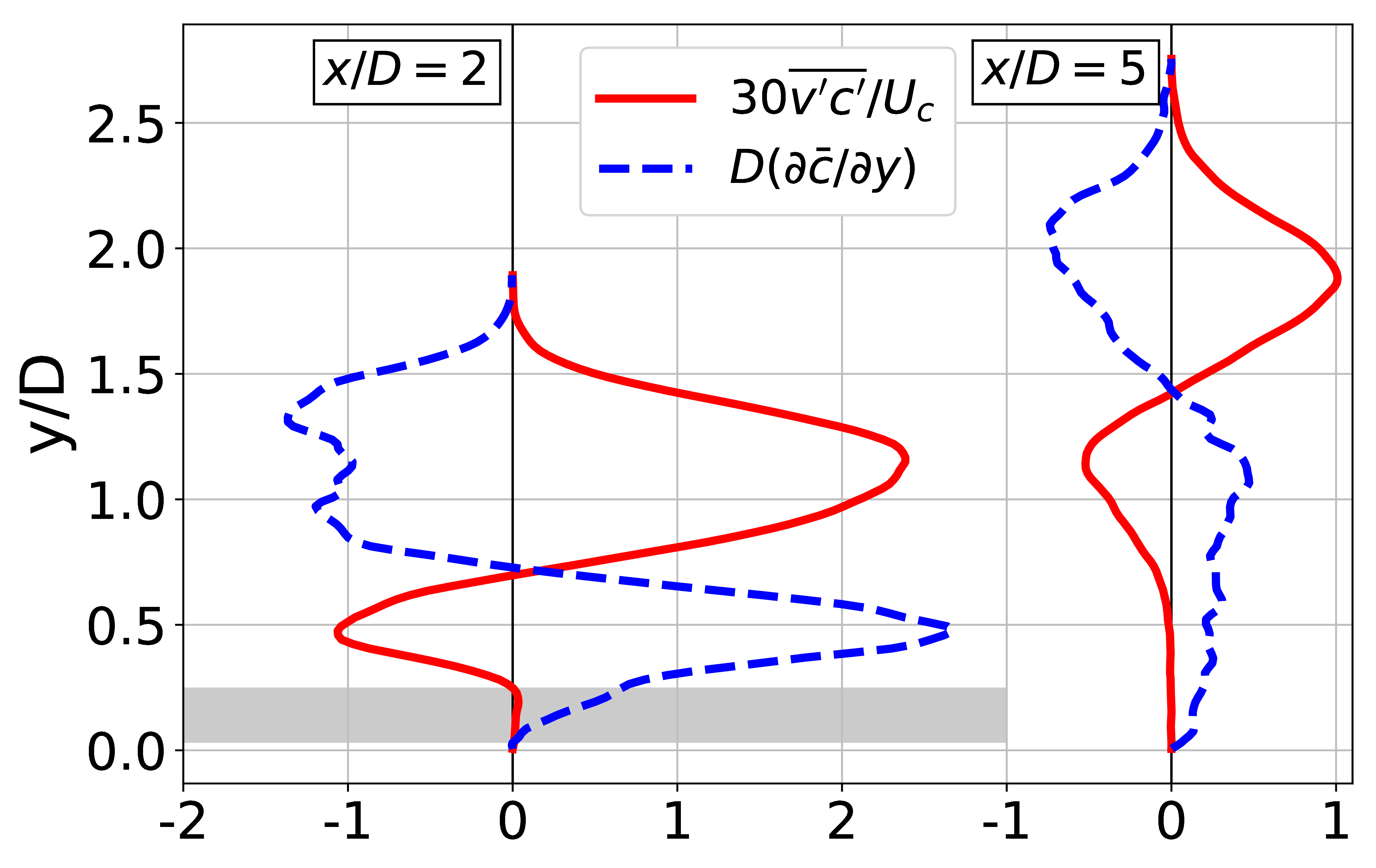} }}%
  \end{center}
\caption{Vertical profiles at the centerplane of the y-component of scalar gradient and turbulent scalar flux. The grey band indicates regions where both have the same sign; for $r=2$ and $x/D=5$, no consistent region of counter gradient transport is observed (thus no grey band), but $\overline{v'c'}$ is very near zero close to the wall.}
\label{fig-6-vc_dcdy}
\end{figure}

A second region where counter gradient transport is present is near the wall, right after injection. In this case, negative diffusivity is observed in the vertical component: both $\overline{v'c'}$ and $\partial \bar{c} / \partial y$ are positive there. Figures~\ref{fig-6-vc_dcdy}(a)-(b) show vertical profiles at $x/D=2$ and $x/D=5$ for both $r=1$ and $r=2$, with grey bands indicating counter gradient transport regions. In general, it seems that the simple GDH could be useful: the solid lines approximately track the negated dashed lines throughout most of the plot, so a moderately non-uniform diffusivity could model $\overline{v'c'}$. However, this pattern breaks down close to the wall in all cases, where $\overline{v'c'}$ either has the same sign as $\partial \bar{c} / \partial y$ or is much smaller in magnitude than expected. 

\begin{figure}
  \begin{center}
    \subfloat[$r=1$, $x/D=2$] {{\includegraphics[width=58mm]{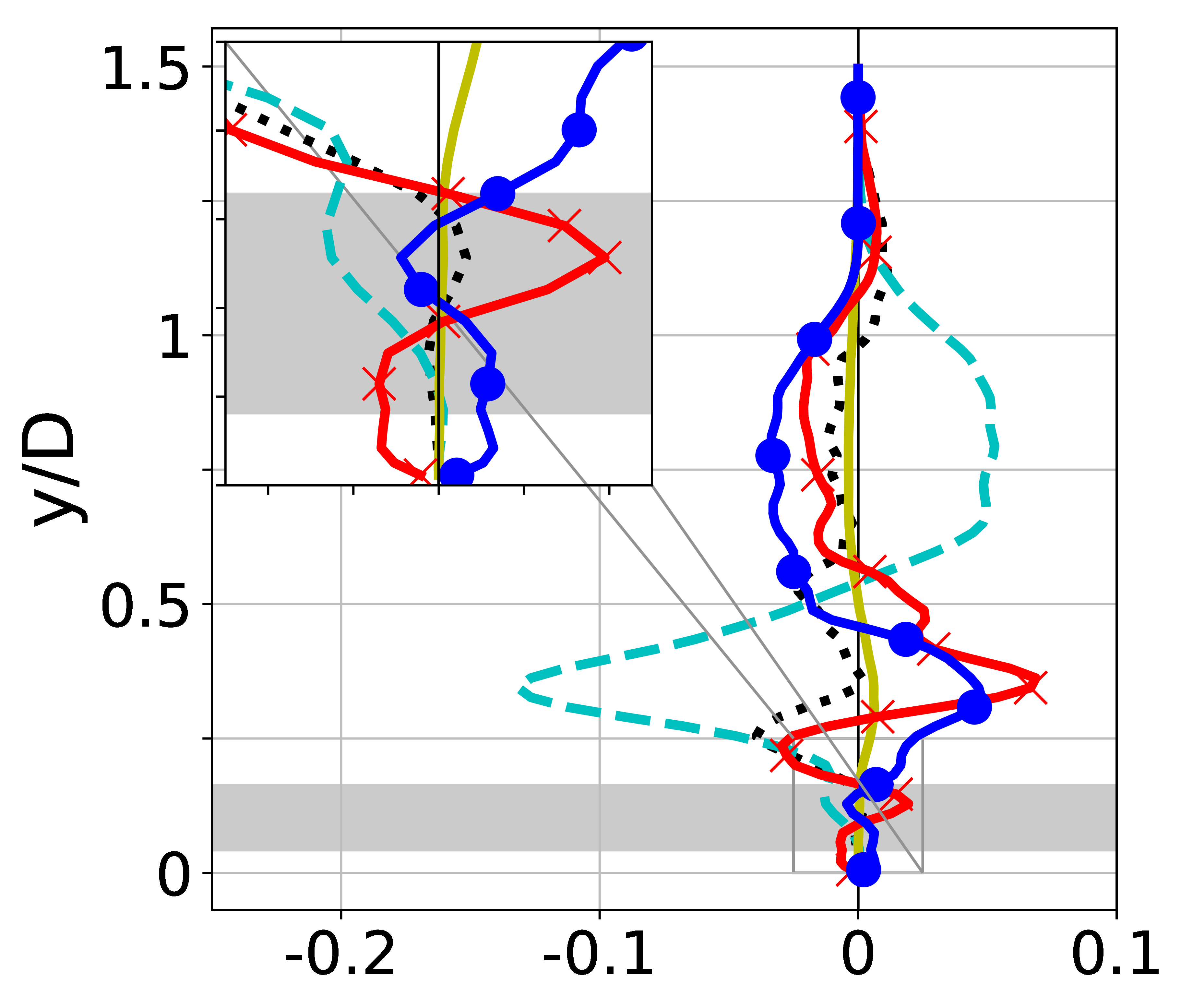} }}%
    \subfloat[$r=1$, $x/D=5$] {{\includegraphics[width=58mm]{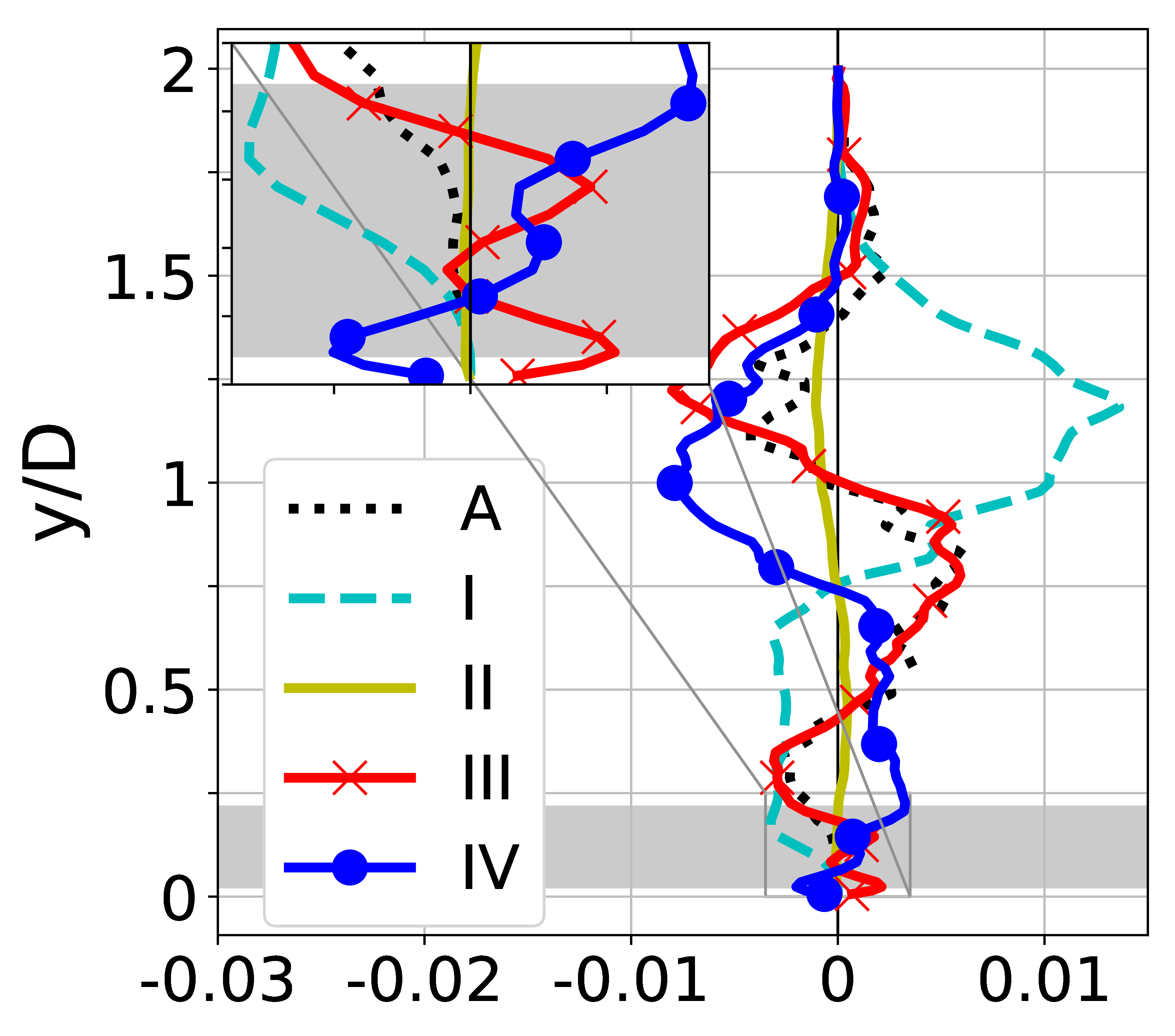} }}\\
    \subfloat[$r=2$, $x/D=2$] {{\includegraphics[width=58mm]{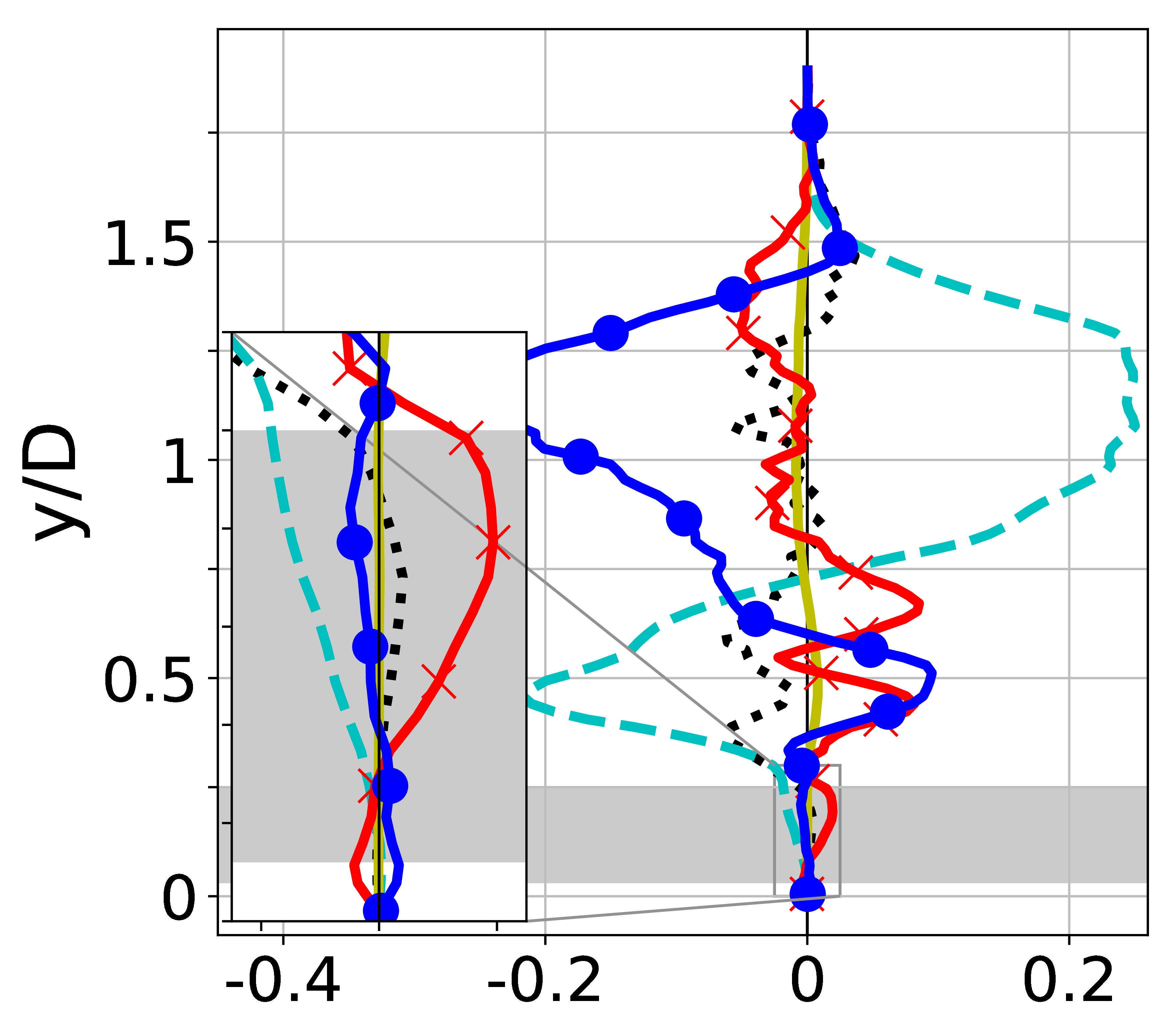} }}%
    \subfloat[$r=2$, $x/D=5$] {{\includegraphics[width=58mm]{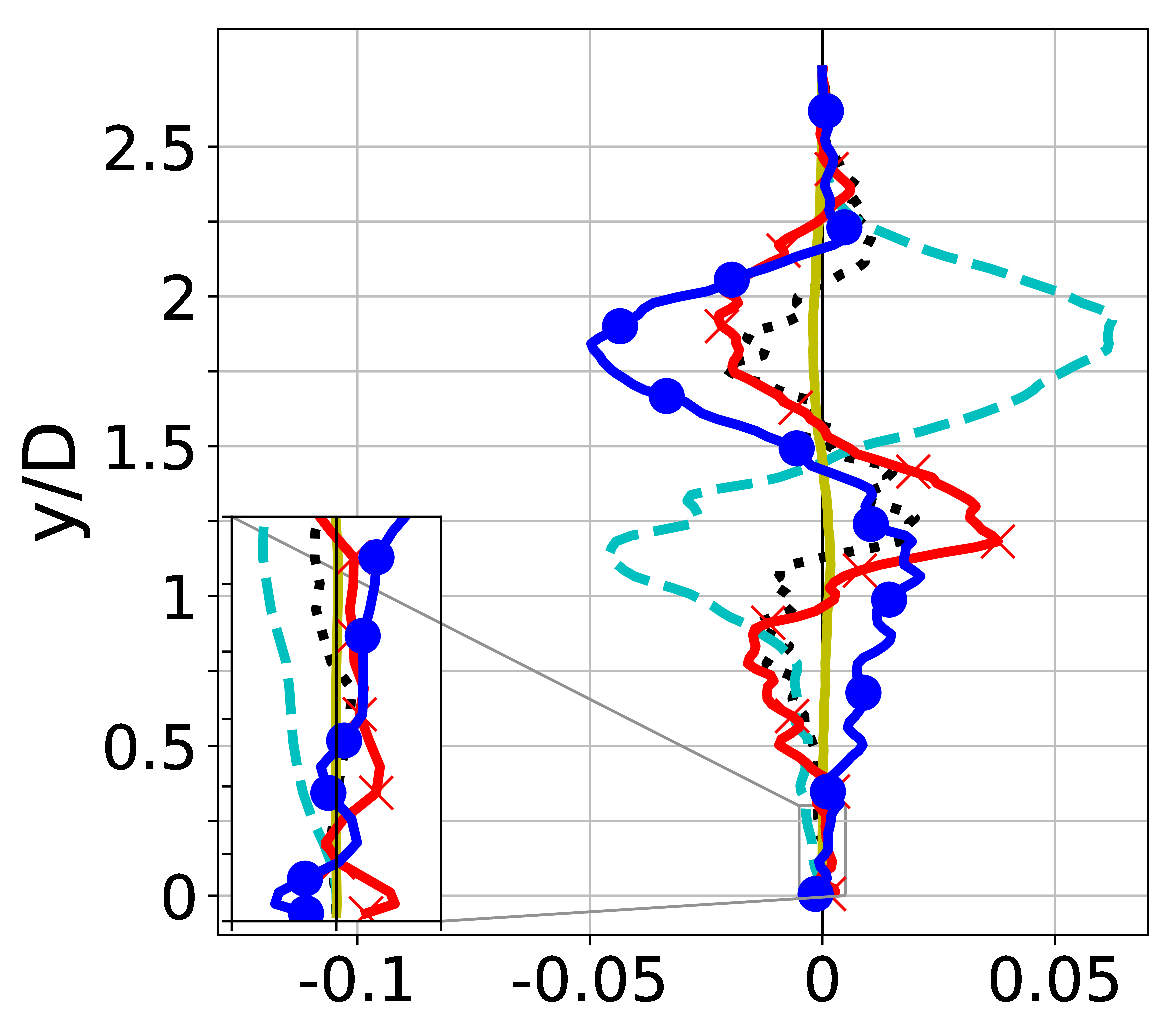} }}%
  \end{center}
\caption{Terms from eq.~\ref{eq-uctransport} in the channel centerline, non-dimensionalized using $D$ and $U_c$. For $r=1$, the zoomed in version shows $0 < y/D < 0.25$. For $r=2$, it shows $0 < y/D < 0.3$. Grey band indicates counter gradient transport in vertical direction.}
\label{fig-7-vc_budget}
\end{figure}

The first hypothesis to explain this instance of counter gradient transport is that it is similar in nature to the one described before, i.e. that gradients in directions orthogonal to y cause a positive correlation between $v'$ and $c'$. However, this is not the case. Averages that are conditional on $u'$ or $w'$ are inconclusive, and the $\overline{v'c'}$ production term is dominated by mean gradients in the y direction. So, the negative diffusivity observed in figure~\ref{fig-6-vc_dcdy} is caused by a different physical mechanism. To help explain it, figure~\ref{fig-7-vc_budget} shows a complete budget of $\overline{v'c'}$, with vertical profiles of all terms of eq.~\ref{eq-uctransport}.

As the plots show, the production term is consistently negative in the grey band, which means that local effects act as a sink and therefore favor a negative value of $\overline{v'c'}$. Since the resulting scalar flux is positive, other effects are overwhelming the production. In particular, close to injection ($x/D=2$), term III is the most positive term. This is the turbulent transport of $\overline{v'c'}$, which unlike the production is inherently non-local. The breakdown of term III (not shown) shows that the molecular terms are negligible throughout the domain, and close to the wall the pressure-scalar correlation is dominant. This suggests that non-local turbulent effects, chiefly through fluctuating pressure, are responsible for generating a positive correlation between $v'$ and $c'$ in this region. Physically, this could be explained by the large scale stirring mentioned by \citet{bodart2013highfidelity}: turbulent eddies with much larger length scales than the length scales over which $\partial \bar{c} / \partial {y}$ varies act in those regions, meaning that they can induce turbulent fluctuations that cannot be explained by local information. Note, however, that \citet{bodart2013highfidelity} pointed out locations of \textit{Type 1} counter gradient transport in their data but provided a physical reasoning that is more appropriate for \textit{Type 2}.

Another interesting observation is on the role of mean flow advection (term A) in the balance. Since A is on the left hand side of eq.~\ref{eq-uctransport}, a positive value represents a net outflow and a negative value represents a net inflow. Even though term A is not as important as some other terms in the grey band, it seems to enter the balance as a net outflow at $x/D=2$ and as a net inflow at $x/D=5$ for both velocity ratios. Since this is mostly due to mean streamwise advection, it suggests that non-local effects generate a positive $\overline{v'c'}$ close to the wall right after injection, which then is advected downstream. In other words, those non-local effects are not as strong at $x/D=5$, but $\overline{v'c'}$ remains positive there partly because it is being advected from upstream (i.e. due to memory effects). This explanation is consistent with figure~\ref{fig-6-vc_dcdy}: the positive $\overline{v'c'}$ close to the wall is more intense and concentrated at $x/D=2$, and it is more diffuse at $x/D=5$.

\subsection{3-dimensional perspective}

In the previous subsections, we highlighted locations where a particular component of the turbulent scalar flux had the same sign as the equivalent component of the mean scalar gradient. We did so for two reasons: first, that is how many previous authors reported counter gradient transport in this flow. For example, \citet{muppidi2008direct}, \citet{bodart2013highfidelity}, and \citet{schreivogel2016simultaneous} noted regions where $\overline{u'c'}$ and $\partial \bar{c} / \partial x$ have the same sign, while \citet{salewski2008mixing} and \citet{milani2018magnetic} highlighted regions where $\overline{v'c'}$ and $\partial \bar{c} / \partial y$ have the same sign; their results are consistent with the locations and mechanisms we explained as \textit{Type 1} and \textit{Type 2} respectively. Second, starting with a 1D perspective permits an easier visualization of the turbulent scalar flux transport equation budgets as done in figures~\ref{fig-5-production_uc} and \ref{fig-7-vc_budget}. However, despite providing useful insights, that methodology is dependent on the particular choice of axis; in fact, unless the mean scalar gradient and the negative turbulent scalar flux vectors are perfectly aligned, it is possible to construct a coordinate frame where one of the components will indicate apparent counter gradient transport. So, in this subsection, we study the same locations by considering the angle between the vectors $-\overline{u_i'c'}$ and $\frac{\partial \bar{c}}{\partial x_i}$, given by $\theta$ in eq.~\ref{eq-angle}. This measure takes all of the 3D information into account and is independent of the choice of axis. In places where the simple GDH of eq.~\ref{eq-gdh} is valid exactly, we should have $\theta = 0$; in places of ``true" counter gradient transport, $\theta > 90^\circ$.

\begin{equation}
\label{eq-angle}
\theta = \arccos{\left(\frac{-\overline{u_i'c'} \frac{\partial \bar{c}}{\partial x_i} }{\sqrt{\overline{u_i'c'} \ \overline{u_i'c'}} \sqrt{\frac{\partial \bar{c}}{\partial x_i} \frac{\partial \bar{c}}{\partial x_i}}}\right)}
\end{equation}

\begin{figure}
  \begin{center}
  \includegraphics[width = 135mm]{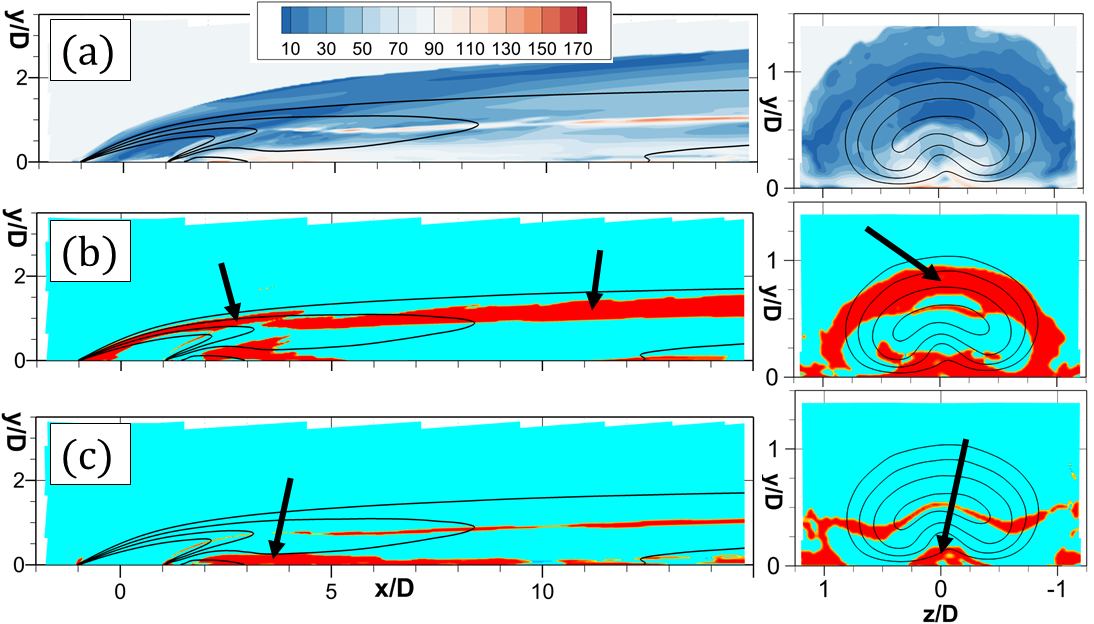}
  \end{center}
\caption{Counter gradient transport in the $r=1$ case. Panels on the left show centerplanes ($z=0$) and on the right show axial planes at $x/D=2$. Lines indicate isocontours of $\bar{c}=0.2, 0.4, 0.6, 0.8$. (a) contains color contours of $\theta$ in degrees. (b) shows, in red, places where $\overline{u'c'} \frac{\partial \bar{c}}{\partial x} > 0$ (with arrows highlighting \textit{Type 1} transport). (c) identifies regions where $\overline{v'c'} \frac{\partial \bar{c}}{\partial y} > 0$ (with arrows highlighting \textit{Type 2} transport).}
\label{fig-8-theta_r2}
\end{figure}

\begin{figure}
  \begin{center}
  \includegraphics[width = 135mm]{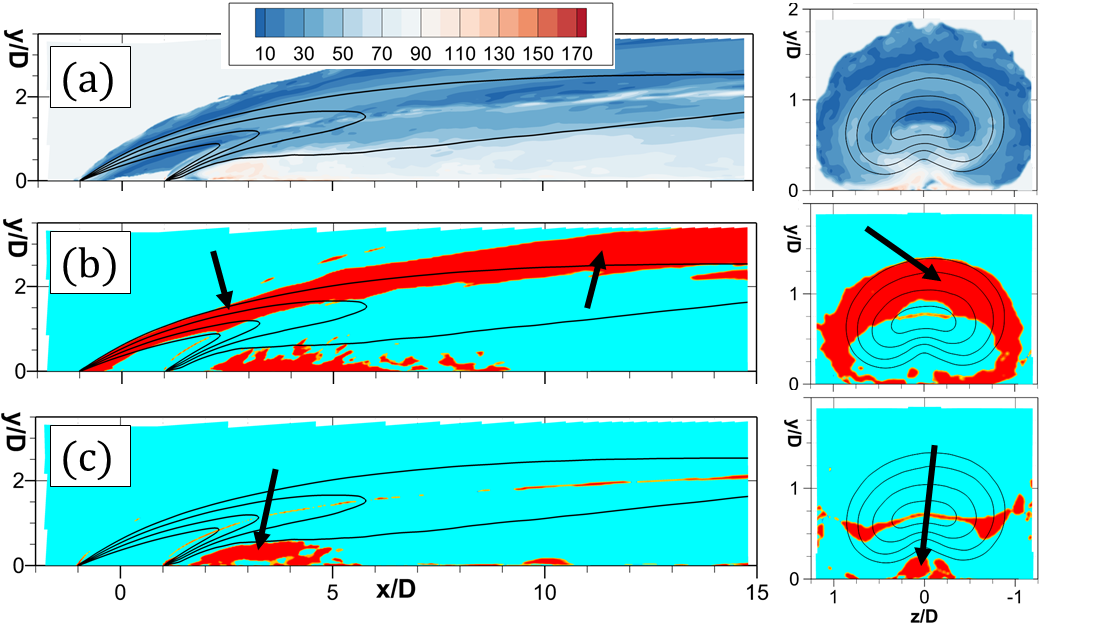}
  \end{center}
\caption{Counter gradient transport in the $r=2$ case. Same description as figure~\ref{fig-8-theta_r2}.}
\label{fig-9-theta_r2}
\end{figure}

Figure~\ref{fig-8-theta_r2}(a) shows color contours of the angle $\theta$ in the $r=1$ LES. Figures~\ref{fig-8-theta_r2}(b)-(c) contain the same planes showing the regions of \textit{Type 1} and \textit{Type 2} regions respectively, as described previously. In general, the angle $\theta$ is highest close to the wall, whose presence induces strong anisotropy into the turbulent mixing. Interestingly, there are key differences between \textit{Type 1} and \textit{Type 2} regions identified previously. On the windward shear layer, where cross-gradient effects give rise to \textit{Type 1} transport, the overall misalignment indicated by $\theta$ is not too high, with angles between $20^\circ$ and $50^\circ$ being common. On the other hand, regions of \textit{Type 2} transport shown in figure~\ref{fig-8-theta_r2}(c) coincide with severe misalignment of $-\overline{u_i'c'}$ and $\frac{\partial \bar{c}}{\partial x_i}$, and in such areas the angle $\theta$ is frequently higher than $90^\circ$. This shows that \textit{Type 1} transport, caused mainly by local effects, actually translates to a consistent but small misalignment between the vectors when seen in 3D. The non-equilibrium, non-local causes of \textit{Type 2} transport actually lead to a more than $90^\circ$ misalignment between the flux and gradient vectors in 3D, which can be seen as a ``true" counter gradient transport that is independent of coordinate frame. Figure~\ref{fig-9-theta_r2}, which presents the same plots for the $r=2$ dataset, leads to similar conclusions.

From a modelling perspective, regions where $\theta$ is significantly above zero cannot be well captured with the simple GDH of eq.~\ref{eq-gdh} irrespective of the diffusivity chosen. Moving to a tensor diffusivity $D_{ij}$, as shown in eq.~\ref{eq-gdhtensor}, can capture cross-gradient effects (since turbulent transport in the $x$ direction can be a function of the $y$ gradient) and would thus be expected to model well the turbulent mixing in regions of \textit{Type 1}. This is further supported by the fact that $\theta$ in those regions is less than $90^\circ$, so the matrix diffusivity $D_{ij}$ is only required to rotate the concentration gradient vector by an acute angle. Thus a positive semi-definite matrix $D_{ij}$ can be used, guaranteeing numerical stability. In locations where \textit{Type 2} transport is found, the non-local effects on $\overline{u_i'c'}$ preclude a purely local model, such as eq.~\ref{eq-gdhtensor}, from capturing all the relevant physics. Even if a diffusivity $D_{ij}$ where chosen such that $\overline{u_i'c'}$ is matched, that matrix would not be positive semi-definite since $\theta > 90^\circ$ in those areas. Therefore, the equation would be numerically unstable, akin to using a negative value of $\alpha_t$ in eq.~\ref{eq-gdh}. For locations of \textit{Type 2} counter gradient transport, only a non-local model, based potentially on solving separate transport equations for $\overline{u_i'c'}$, could hope to reproduce the appropriate physics.

\section{Deep learning modelling}

In this section, we transition to the question of modelling the $\overline{u_i'c'}$ vector using deep learning. For simplicity, this section makes heavy use of matrix notation instead of index notation: a lower case bold letter indicates a vector (first order tensor) and an upper case bold letter indicates a 3x3 matrix (second order tensor).

\subsection{Tensor basis neural network for scalar flux modelling}

Given the promise of the tensor basis neural network (TBNN) proposed by \citet{julia_deepnn}, we would like to use a similar method to model the vector $\overline{\boldsymbol{u}'c'}$. In summary, we aim to employ a neural network architecture that is designed to respect rotational invariance while predicting a tensorial output. However, the architecture described in \citet{julia_deepnn} is designed to prescribe the Reynolds stress anisotropy tensor $\mathsfbi{B}$, which is a symmetric and traceless second order tensor. The approach must be modified for the turbulent scalar flux vector $\overline{\boldsymbol{u}'c'}$, which does not have those characteristics.

We start by assuming that the turbulent scalar flux is a general vector-valued function of the mean velocity and scalar gradients, $\nabla \bar{\boldsymbol{u}}$ and $\nabla \bar{c}$. The velocity gradient tensor is split into a symmetric and an anti-symmetric part, $\mathsfbi{S}$ and $\mathsfbi{R}$ respectively, such that $\nabla \bar{\boldsymbol{u}} = \mathsfbi{S} + \mathsfbi{R}$. As is done in previous work that models turbulent scalar fluxes \citep{milani2018approach}, two additional dimensionless quantities are also used to regress $\overline{\boldsymbol{u}'c'}$: the Reynolds number based on wall distance, $Re_d = \sqrt{k}d/\nu$, and the eddy viscosity ratio, $\nu_t / \nu$, where $d$ is the distance to the nearest wall, and $k$ is the turbulent kinetic energy, and $\nu_t$ is the eddy viscosity from a baseline turbulence model. So, the functional dependence is:

\begin{equation}
\label{eq-function}
-\overline{\boldsymbol{u}'c'} = \boldsymbol{f}(\mathsfbi{S}, \mathsfbi{R}, \nabla \bar{c}, \nu_t / \nu, Re_d).
\end{equation}

Note that the model in eq.~\ref{eq-function} depends on a baseline turbulence model for the momentum equations, which provides field quantities such as $k$ and $\nu_t$. This is a design choice that makes the model more usable in practice, since one would not have access to high fidelity simulation values for these quantities in a practical RANS simulation. A similar approach was taken by other authors who worked in scalar flux modelling with machine learning techniques \citep[e.g.][]{sandberg2018applying, sotgiu2018turbulent}. As is done in \citet{julia_deepnn}, eq.~\ref{eq-function} is cast as a summation of an appropriate tensor basis, where each basis element is a vector and is multiplied by a scalar factor $g^{(n)}$. This quantity, in turn, can be an arbitrary function of invariants $\lambda_j$, which are scalar quantities derived from the tensors $\mathsfbi{S}$, $\mathsfbi{R}$ and $\nabla \bar{c}$ that are reference frame independent. Thus, $\overline{\boldsymbol{u}'c'}$ becomes:

\begin{equation}
\label{eq-basissum}
-\overline{\boldsymbol{u}'c'} = \sum_{n=1}^{6}g^{(n)}(\lambda_1, \lambda_2, ..., \lambda_{15}) \boldsymbol{t}^{(n)}.
\end{equation}

\noindent When the function is written in this form, it is guaranteed to be unchanged under any coordinate frame rotation and reflection as explained in \citet{julia_deepnn}. The neural network's eventual goal will be to approximate the functions $g^{(n)}(\lambda_1, \lambda_2, ..., \lambda_{15})$.

For this problem, we cannot use the same basis and invariants used in \citet{pope} or \citet{julia_deepnn}, since the quantity of interest is a vector ($\overline{\boldsymbol{u}'c'}$) that depends upon one symmetric matrix ($\mathsfbi{S}$), one anti-symmetric matrix ($\mathsfbi{R}$), and one vector ($\nabla \bar{c}$). Turning to the review of \citet{zheng1994theory}, one can find the appropriate vector basis, given as

\begin{align}
\begin{split}
\label{eq-tensorbasis}
&\boldsymbol{t}^{(1)} = \nabla \bar{c}, \quad \boldsymbol{t}^{(2)} = \mathsfbi{S} \nabla \bar{c}, \quad \boldsymbol{t}^{(3)} = \mathsfbi{R} \nabla \bar{c} \\
&\boldsymbol{t}^{(4)} = \mathsfbi{S}^2 \nabla \bar{c}, \quad \boldsymbol{t}^{(5)} = \mathsfbi{R}^2 \nabla \bar{c}, \quad \boldsymbol{t}^{(6)} = \left( \mathsfbi{S}\mathsfbi{R} + \mathsfbi{R}\mathsfbi{S} \right)\nabla \bar{c}
\end{split}
\end{align}

\noindent and the appropriate invariants for the incompressible case, given as

\begin{align}
\begin{split}
\label{eq-invariants}
&\lambda_1 = tr(\mathsfbi{S}^2), \quad \lambda_2 = tr(\mathsfbi{S}^3), \quad \lambda_3 = tr(\mathsfbi{R}^2), \quad \lambda_4 = tr(\mathsfbi{S}\mathsfbi{R}^2), \quad \lambda_5 = tr(\mathsfbi{S}^2\mathsfbi{R}^2)\\
&\lambda_6 = tr(\mathsfbi{S}^2\mathsfbi{R}^2\mathsfbi{S}\mathsfbi{R}) , \quad \lambda_7 = \nabla \bar{c}^T\nabla \bar{c}, \quad \lambda_8 = \nabla \bar{c}^T\mathsfbi{S}\nabla \bar{c}, \quad \lambda_9 = \nabla \bar{c}^T\mathsfbi{S}^2\nabla \bar{c}\\
&\lambda_{10} = \nabla \bar{c}^T\mathsfbi{R}^2\nabla \bar{c}, \quad \lambda_{11} = \nabla \bar{c}^T\mathsfbi{S}\mathsfbi{R}\nabla \bar{c}, \quad \lambda_{12} = \nabla \bar{c}^T\mathsfbi{S}^2\mathsfbi{R}\nabla \bar{c} \\ 
&\lambda_{13} = \nabla \bar{c}^T\mathsfbi{R}\mathsfbi{S}\mathsfbi{R}^2\nabla \bar{c}, \quad \lambda_{14} = Re_d, \quad \lambda_{15} = \nu_t / \nu.
\end{split}
\end{align}

\noindent Note that the first 13 invariants are obtained from the tensor inputs, and the last two are purely the scalar quantities $Re_d$ and $\nu_t / \nu$ (which are automatically invariant to the choice of coordinate frame). In eqs.~\ref{eq-tensorbasis} and \ref{eq-invariants}, products denote standard matrix-matrix or matrix-vector products, and $tr()$ denotes the trace of a matrix.

Under this formulation, given a vector basis and the invariants, the neural network directly prescribes the vector $\overline{\boldsymbol{u}'c'}$ whose divergence acts as a source term in the Reynolds averaged scalar transport equation. As discussed in \citet{wu2019reynolds}, such an approach can lead to an ill conditioned equation, which is difficult to solve numerically. To mitigate that, they suggest that whenever possible data-driven closures should aim to predict diffusivity coefficients instead of source terms. With that in mind, we note that all basis elements shown in eq.~\ref{eq-tensorbasis} consist of a matrix multiplying the mean concentration gradient. So, we can define a different basis $\mathsfbi{T}^{(n)}$ where $\boldsymbol{t}^{(n)} = \mathsfbi{T}^{(n)} \nabla \bar{c}$. Thus, we rewrite eq.~\ref{eq-basissum} as:

\begin{equation}
\label{eq-basissum_2}
-\overline{\boldsymbol{u}'c'} = \overbrace{\left[ \sum_{n=1}^{6}g^{(n)}(\lambda_1, \lambda_2, ..., \lambda_{15}) \mathsfbi{T}^{(n)} \right]}^\mathsfbi{D} \nabla \bar{c},
\end{equation}

\noindent with the matrix part of the basis given by:

\begin{align}
\begin{split}
\label{eq-tensorbasis_2}
&\mathsfbi{T}^{(1)} = \mathsfbi{I}, \quad \mathsfbi{T}^{(2)} = \mathsfbi{S}, \quad \mathsfbi{T}^{(3)} = \mathsfbi{R}\\
&\mathsfbi{T}^{(4)} = \mathsfbi{S}^2, \quad \mathsfbi{T}^{(5)} = \mathsfbi{R}^2, \quad \mathsfbi{T}^{(6)} = \mathsfbi{S}\mathsfbi{R} + \mathsfbi{R}\mathsfbi{S},
\end{split}
\end{align}

\noindent where $\mathsfbi{I}$ is the 3x3 identity matrix.

In eq.~\ref{eq-basissum_2}, the term in square brackets is naturally interpreted as a 3x3 turbulent diffusivity matrix $\mathsfbi{D}$. Thus, we recover an equation of the same form as eq.~\ref{eq-gdhtensor} that is amenable to be learned via a neural network with embedded rotational invariance.

Before the full model architecture is discussed, there remains the issue of non-dimensionalization. To obtain relevant scales and values for the eddy viscosity $\nu_t$, a baseline turbulence model is run on the original mesh utilizing the frozen mean velocity field from the LES calculation. Any model that yields length and time scales and a value for the eddy viscosity could be employed with the current framework; in the present paper, we use the realizable $k-\epsilon$ model of \citet{shih_realizable}. So, given the mean LES velocity and pressure fields, the transport equations for $k$ and $\epsilon$ are solved, which also generates $\nu_t$ as a by-product. This approach to non-dimensionalization has been used by others such as \citet{sandberg2018applying} and \citet{milani2020generalization}. Then, $\mathsfbi{S}$, $\mathsfbi{R}$, and $\nabla \bar{c}$ in equations eq.~\ref{eq-invariants} and eq.~\ref{eq-tensorbasis_2} are non-dimensionalized using local values of $k$ and $\epsilon$ before being fed as inputs to the network. To guarantee that the neural network maps from dimensionless inputs to a dimensionless output, a non-dimensional diffusivity tensor $\mathsfbi{D}^*$ is explicitly introduced as $\mathsfbi{D} = \nu_t \mathsfbi{D}^*$.

\begin{figure}
  \begin{center}
  \includegraphics[width = 100mm]{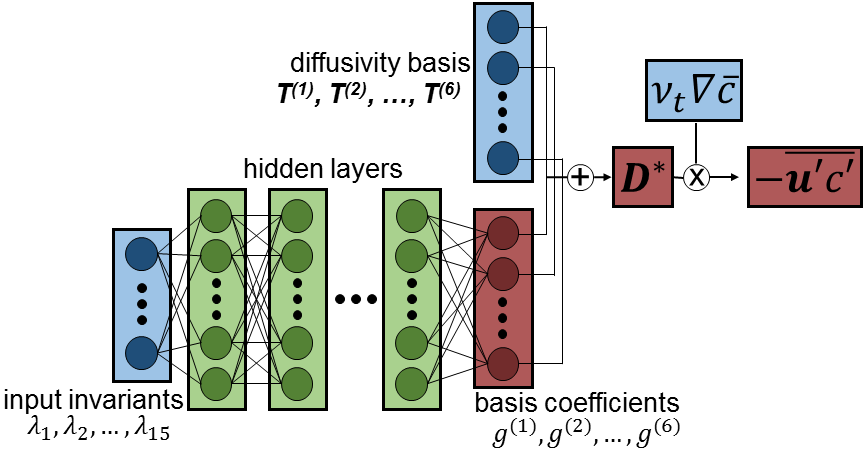}
  \end{center}
\caption{Schematic of TBNN-s architecture. Blue cells represent inputs to the network, green cells represent the hidden layers containing tunable parameters, and red cells denote different levels of outputs.}
\label{fig-10-tbnns}
\end{figure}

The architecture proposed, henceforth referred to as tensor basis neural network for scalar flux modeling, or TBNN-s, is presented in figure~\ref{fig-10-tbnns}. The network receives as inputs the mean flow invariants (listed in eq.~\ref{eq-invariants}) non-dimensionalized by $k$ and $\epsilon$ and processes them through a sequence of hidden layers. These are fully connected layers, which sequentially perform a linear transformation (matrix multiplication plus an offset) followed by a non-linear activation function (the simple rectified linear unit, or ReLU, is employed). The last hidden layer does not use a ReLU function and transforms its inputs into a 6-dimensional vector containing the values of $g^{(n)}$, the dimensionless basis coefficients. Note that the hidden layers are responsible for explicitly providing the functions $g^{(n)}(\lambda_1, ..., \lambda_{15})$ and contain all the tunable parameters of the model, which are learned from training data. Then, the coefficients $g^{(n)}$ multiply elementwise their respective dimensionless basis tensor $\mathsfbi{T}^{(n)}$ shown in eq.~\ref{eq-tensorbasis_2} and the result is summed producing the dimensionless diffusivity matrix $\mathsfbi{D}^*$. Finally, the diffusivity is made dimensional via multiplication by $\nu_t$ and its matrix-vector product with the mean scalar gradient produces $-\overline{\boldsymbol{u}'c'}$.

Note that this modelling framework is similar to that of \citet{weatheritt2020data}, who also expanded the turbulent scalar flux vector as a sum of tensor basis that maintains rotational invariance. Our model has two key differences: first, we use a deep neural network to approximate the functions $g^{(n)}(\lambda_i)$, while \citet{weatheritt2020data} use gene expression programming, a different regression tool (see \citet{weatheritt2017comparative} for a discussion of their differences). Second, their basis also involves the Reynolds stress tensor. In theory this allows for more accurate representation of the scalar flux, as they confirm. However, in a typical RANS framework one does not have access to the true values of the Reynolds stress; as such, we preferred to develop models that are based solely on the mean velocity gradient. In future work, it would be useful to compare their model to ours under the same conditions in order to analyze strengths and weaknesses of different data driven turbulence modelling strategies.

\subsection{Loss function and implementation}

Neural networks are trained via gradient descent-like algorithms. A loss function $J$ is defined to capture how inadequate the present solution is compared to the known results from training data. Given multiple batches of training data, the loss function is calculated together with its gradient with respect to each trainable parameter. The gradient, a key component for training neural networks, is calculated using backpropagation, which is sequential application of the chain rule starting at the loss function $J$ and progressing back through the layers all the way to the first hidden layer. For backpropagation to work, every operation defined in the network must be differentiable which is the case for figure~\ref{fig-10-tbnns}. Note that the loss function is defined based on the mismatch between the predicted value of the turbulent scalar flux, $\overline{\boldsymbol{u}'c'}_{PRED}$, and the turbulent scalar flux given in the training data, $\overline{\boldsymbol{u}'c'}_{LES}$. We experimented with a few different options, including the common sum of squared differences and sum of absolute differences:

\begin{equation}
\label{eq-losses_1}
J_{L1} = \frac{1}{N_{data}} \sum_{i=1}^{N_{data}} ||\overline{\boldsymbol{u}'c'}^{(i)}_{PRED} - \overline{\boldsymbol{u}'c'}^{(i)}_{LES} ||_1,
\end{equation}

\begin{equation}
\label{eq-losses_2}
J_{L2} = \frac{1}{N_{data}} \sum_{i=1}^{N_{data}} ||\overline{\boldsymbol{u}'c'}^{(i)}_{PRED} - \overline{\boldsymbol{u}'c'}^{(i)}_{LES} ||^2_2,
\end{equation}

\noindent where $|| \ ||_1$ indicates vector 1-norm, $|| \ ||_2$ indicates vector 2-norm, $N_{data}$ is the number of data points over which the loss is computed, and a superscript $i$ indicates the i-th training point. These definitions of the loss have the disadvantage that they depend on a particular non-dimensionalization and scaling to compare points from different datasets and different locations from the same dataset. Instead, if we desire a loss function whose gradient with respect to model parameters is independent of the scaling used, the mean of the natural logarithms of the vector norms is a possible choice:

\begin{equation}
\label{eq-losses_log}
J_{log} = \frac{1}{N_{data}} \sum_{i=1}^{N_{data}} log \left( \frac{||\overline{\boldsymbol{u}'c'}^{(i)}_{PRED} - \overline{\boldsymbol{u}'c'}^{(i)}_{LES} ||_2}{|| \overline{\boldsymbol{u}'c'}^{(i)}_{LES} ||_2} \right).
\end{equation}

\noindent In eq.~\ref{eq-losses_log}, the factor that scales the loss at each point is the $|| \overline{\boldsymbol{u}'c'}^{(i)}_{LES} ||_2$ in the denominator. Because the logarithm of the ratio becomes a difference of logarithms, this term vanishes when the gradient of $J_{log}$ with respect to the neural network parameters is calculated. This scaling is chosen so that the loss attains readily interpretable values (for example, a loss around 0 indicates that the vectorial discrepancy between predicted and given scalar fluxes has roughly the same norm as the LES scalar flux itself).

We experimented with the three loss definitions shown in eqs.~\ref{eq-losses_1}, \ref{eq-losses_2} and \ref{eq-losses_log}. The results on the mean scalar field $\bar{c}$ were similar among the three, so a comparison will not be shown for brevity. For the results in the present paper, the function $J_{log}$ of eq.~\ref{eq-losses_log} is employed. We also added an L2 regularization term to the loss, which penalizes high magnitude parameters in the model. This is commonly done to prevent overfitting. The loss is optimized with a widely used algorithm based on stochastic gradient descent, called adaptive moment estimation (Adam) from \citet{kingma2014adam}. Part of the reason for the current popularity of this method is that it has been shown to be reasonably robust to the choice of hyperparameters, especially the learning rate.

To choose the different hyperparameters for the TBNN-s, we used validation points that were not used during training. \citet{milani2019enriching} presents data on the geometry of figure~\ref{fig-1-schematicsdomain} for a third value of velocity ratio, $r=1.5$. In the current paper, the TBNN-s is trained exclusively on the $r=2$ dataset and the hyperparameters are chosen such that good values of the loss function are obtained on points taken from the $r=1.5$ case. We observed that the quality of the model generated is not a strong function of the hyperparameters considered within a reasonable range. Besides, the validation aims to find hyperparameters that minimize the loss in eq.~\ref{eq-losses_log}, when the true test for the model is how well it predicts the mean scalar field $\bar{c}$ after the Reynolds averaged equations are solved. Thus, not much emphasis should be placed on the specific choice of hyperparameters, and we picked a set that performed well on the validation set after manual tuning. In the current work, we use a learning rate of $10^{-3}$ and other default parameters for Adam; we employ 10 hidden layers with 30 nodes each, roughly in line with the network size of \citet{julia_deepnn}; we use an L2 regularization strength of $10^{-2}$; we use an additional regularization method called dropout \citep{srivastava2014dropout} with a drop probability of $0.2$; and the batch size in which training data is fed into the Adam optimizer is 50. Training a neural network is not a deterministic exercise because results depend on a random initialization and on the stochastic nature of the optimization algorithm. To mitigate that, several training runs were executed with the same hyperparameters and the network that was saved and employed in this paper is the one that, at any point during each training procedure, produced the minimum loss in the validation set.

The network as shown in figure~\ref{fig-10-tbnns} is implemented in Python using the Tensorflow package \citep{abadi2016tensorflow} and the code can be found online in the following GitHub repository: \url{https://github.com/pmmilani/tbnns}. The TBNN-s discussed in following subsections was trained on 100k randomly sample points from the inclined jet in crossflow with $r=2$. Only points where the mean scalar gradient is significant ($||\nabla \bar{c}||_2 / (\epsilon / k^{1.5}) > 10^{-3}$) are considered at training time. 

\subsection{Results}

The architecture described in Section 4.1 is used in its entirety when the network is being trained and a correct value of $\overline{\boldsymbol{u}'c'}$ is available. Once the network has been trained, it is applied up to the prediction of $\mathsfbi{D}$. In other words, the network learns to predict a turbulent diffusivity tensor that best matches the turbulent scalar flux presented at training time even though it has never seen an input turbulent diffusivity tensor. When applied to a test set, it predicts a $\mathsfbi{D}$ field over the whole domain which is then used to solve the Reynolds averaged scalar equation

\begin{equation}
\label{eq-raad}
\frac{\partial}{\partial x_i} (\bar{u}_i \bar{c}) = \frac{\partial}{\partial x_i} \left( \frac{\nu}{Sc} \frac{\partial \bar{c}}{\partial x_i} \right) + \frac{\partial}{\partial x_i} \left( D_{ij} \frac{\partial \bar{c}}{\partial x_j} \right),
\end{equation}

\noindent where the last term on the right hand side consists of the model for the turbulent scalar flux involving the turbulent diffusivity matrix. To test the model independent of any velocity field errors, eq.~\ref{eq-raad} is solved with the mean velocity field provided by the LES and with the matrix turbulent diffusivity $\mathsfbi{D}$ provided by the TBNN-s. Note that the TBNN-s is run only once, using LES results with $k$ and $\epsilon$ fields from the baseline turbulence model (realizable $k-\epsilon$) as inputs, and then the diffusivity matrix is fixed and eq.~\ref{eq-raad} is solved on the LES mesh. We employ ANSYS Fluent for the baseline turbulence model and for solving the scalar transport equation.

Before discussing the results, there is one important consideration to be made about the diffusivity matrix $\mathsfbi{D}$. An arbitrary scalar (isotropic) diffusivity could be numerically unstable, namely if it were ever negative. Analogously, a matrix diffusivity $\mathsfbi{D}$ produced by the TBNN-s could also lead to an unstable partial differential equation under some conditions, so it must be post-processed before we use it in eq.~\ref{eq-raad}. In particular, we require that the diffusivity $\mathsfbi{D}$ be a positive semi-definite matrix. Note that this is a requirement for stability for any closure of the form given in eq.~\ref{eq-gdhtensor}, including higher order algebraic closures. See appendix~B for a more detailed explanation of this point. Therefore, we post process the diffusivity $\mathsfbi{D^*}$ output: in locations where the predicted matrix is not positive semi-definite, it is replaced by a diagonal matrix that uses the GDH with fixed $Sc_t=0.85$ as the model.

\subsubsection{A priori results}

A priori results directly test the TBNN-s predictions of the scalar flux. All the results shown are for a network trained on the $r=2$ dataset only.

\begin{figure}
  \begin{center}
  \includegraphics[width = 130mm]{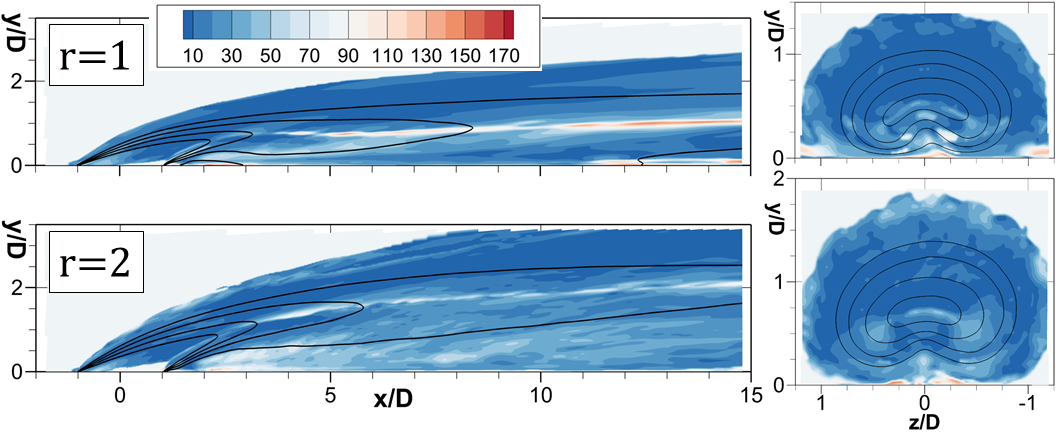}
  \end{center}
\caption{Angle between modelled and true turbulent scalar flux in degrees, given in $\theta^\text{TBNN-s}$. Left shows centerplane ($z=0$) and right shows axial planes at $x/D=2$; black lines indicate isocontours of $\bar{c}=0.2, 0.4, 0.6, 0.8$.}
\label{fig-11-theta_tbnns}
\end{figure}

Figure~\ref{fig-11-theta_tbnns} shows the alignment between the predicted turbulent scalar flux vector and the ground truth scalar flux vector from the LES for both datasets. The angle in degrees is calculated as

\begin{equation}
\label{eq-angle_tbnns}
\theta^\text{TBNN-s} = \arccos{\left(\frac{-\overline{u_i'c'} \left( D_{ij} \frac{\partial \bar{c}}{\partial x_j} \right) }{\sqrt{\overline{u_i'c'} \ \overline{u_i'c'}} \sqrt{\left( D_{ij} \frac{\partial \bar{c}}{\partial x_j} \right) \left( D_{ij} \frac{\partial \bar{c}}{\partial x_j} \right)}}\right)}
\end{equation}

\noindent where $D_{ij}$ is the matrix turbulent diffusivity predicted by the TBNN-s.

Note that these plots should be compared to the misalignment that would arise from a baseline GDH model shown in Section 3 in figures~\ref{fig-8-theta_r2}(a) and \ref{fig-9-theta_r2}(a). In figure~\ref{fig-11-theta_tbnns}, we see significant improvement in the full domain for both cases, $r=2$ (where the model was trained) and $r=1$ (where the model's generality is tested). In the windward shear layer, where \textit{Type 1} counter gradient transport was observed, the alignment is now excellent, with values below $10^\circ$ being prevalent. This supports the argument made in Section 3 that this physical process would be amenable to be modelled with a local, but anisotropic model. The alignment near the wall is also significantly superior to that shown in figures~\ref{fig-8-theta_r2}(a) and \ref{fig-9-theta_r2}(a). However, some higher values of $\theta^\text{TBNN-s}$ are still present there. First, this reflects the fact that whenever the original misalignment, between $-\overline{u_i'c'}$ and $\frac{\partial \bar{c}}{\partial x_j}$, is more than $90^\circ$ it cannot be fully fixed with a positive semi-definite diffusivity matrix. Second, as argued in Section 3, the turbulent scalar flux in this near-wall region is affected by non-equilibrium, non-local effects which cannot be satisfactorily captured with a purely local model.

\begin{figure}
  \begin{center}
  \includegraphics[width = 130mm]{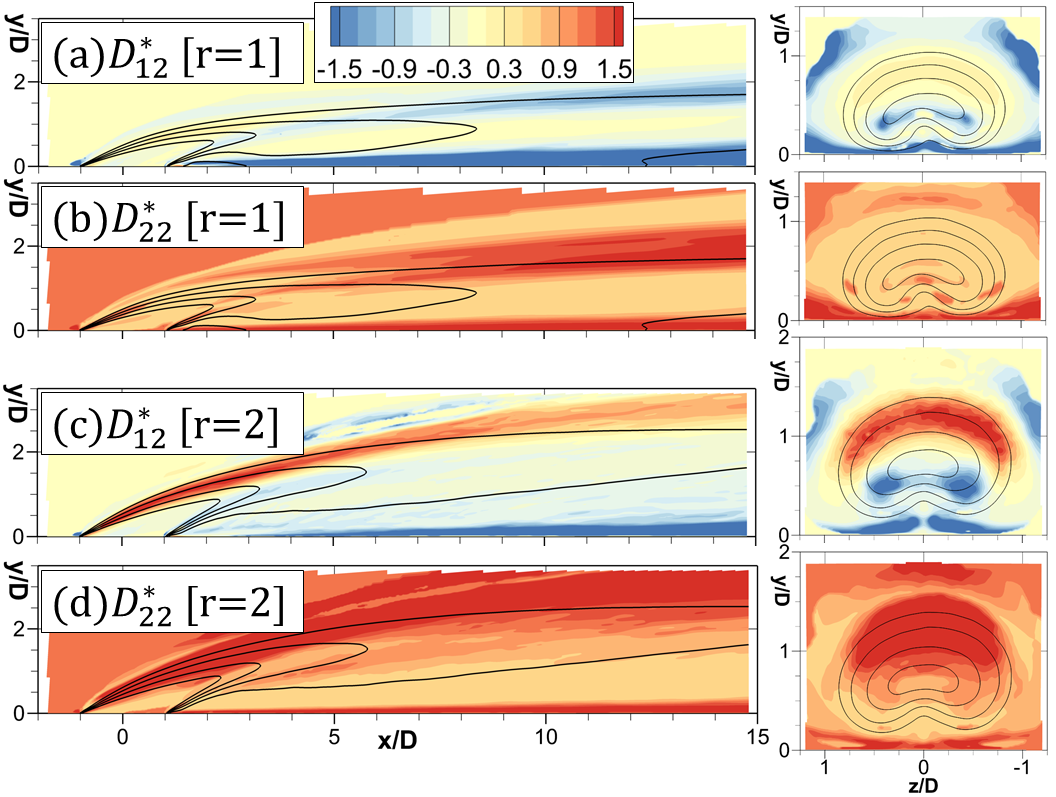}
  \end{center}
\caption{Dimensionless components of the diffusivity matrix $\mathsfbi{D}$ predicted by the TBNN-s in $r=1$ and $r=2$ cases. Left shows centerplane ($z=0$) and right shows axial planes at $x/D=2$; black lines indicate isocontours of $\bar{c}=0.2, 0.4, 0.6, 0.8$.}
\label{fig-12-D_tbnns}
\end{figure}

Figure~\ref{fig-12-D_tbnns} shows two different components of the turbulent diffusivity matrix predicted by the TBNN-s in the $r=1$ and $r=2$ cases. $D^*_{12}$ is an off-diagonal component that controls the effect of wall-normal gradients on the streamwise transport (and thus is relevant for prediction of \textit{Type 1} transport as described in Section 3), and $D^*_{22}$ is the wall-normal diagonal element. Note that a standard GDH model would have uniform components, with $D^*_{12}$ being zero and $D^*_{22}$ being $1/Sc_t$.

The component $D^*_{12}$ has low magnitude in most regions of the flow, showing that cross-gradients effects are not important everywhere. But it is particularly active where the wall-normal mean velocity gradient is significant. This is to be expected, since strong vertical gradients in $\bar{u}$ would cause vertical eddies to correlate $u'$ with $c'$ in the presence of vertical concentration gradients. Close to the wall, this term is largest in magnitude because of the high velocity gradient due to the no-slip condition. It is also active in the windward shear layer, in locations of \textit{Type 1} transport. While interesting from a physical perspective, these improvements by themselves do not lead to a significant improvement in mean scalar concentration, as discussed in appendix~A. Note that the TBNN-s prediction of $D^*_{12}$ in the windward shear layer has different signs in the $r=1$ and $r=2$ cases, which is the correct behavior since $\partial \bar{u} / \partial y$ is positive for $r=1$ and negative for $r=2$ in that region. This showcases the capacity of the model to learn physics rather than just regurgitate the training data; in this case, this is due to the basis expansion in eq.~\ref{eq-basissum_2}, which explicitly multiplies the data driven functions, $g^{(i)}$, by terms such as $\mathsfbi{S}$ which are sensitized to the local mean flow gradients.

The component $D^*_{22}$ is positive everywhere as it should be, due to the positive semi-definite requirement on $\mathsfbi{D}$. It does contain some spatial variation. In general, it is lower in the bottom half on the jet and higher in the top half, except for very near the wall where it attains the highest values. These results are consistent with the experimental results of \citet{kohli2005turbulent} who reported smaller vertical heat diffusivity below the jet's centreline compared to above it. Relatively high values of $D^*_{22}$ right above the wall and after injection appear in both the training set and in the test set, which is inconsistent with the regions of \textit{Type 2} transport. This is because from the argument in section 3.2, we would expect $D^*_{22}$ to be physically very small or even negative in this region since the counter gradient turbulent transport in the wall normal direction is not caused by cross-gradients in the mean concentration. A possible reason for this inconsistency is that in those regions both the turbulent transport and the mean scalar gradients have very low magnitude, so the diffusivity entries are, in a sense, a ratio between two very small numbers. Thus, the TBNN-s which attempts to predict this ratio might fail to obtain satisfactory results.

\subsubsection{A posteriori results}

\begin{figure}
  \begin{center}
  \includegraphics[width = 130mm]{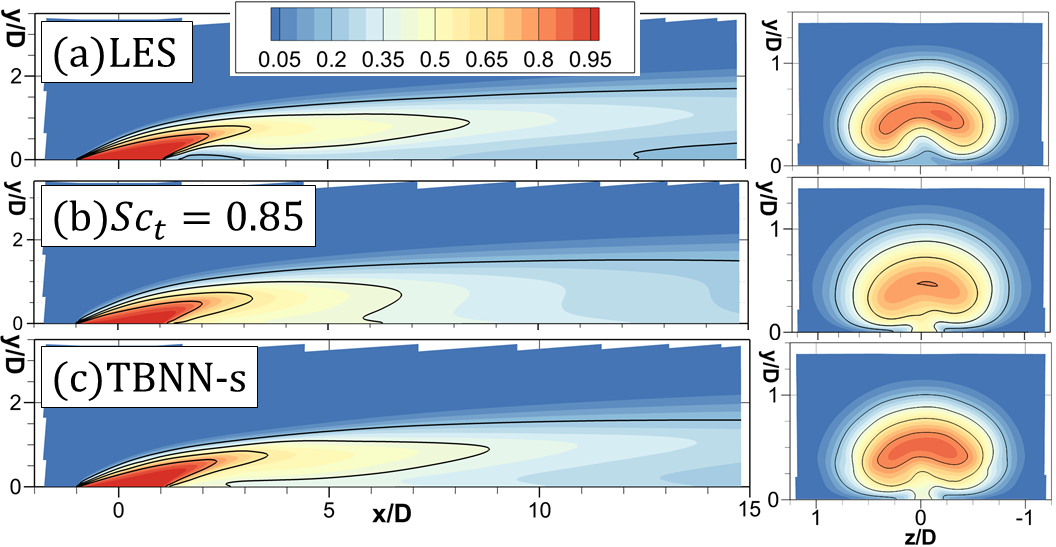}
  \end{center}
\caption{Mean scalar concentration contours in the $r=1$ case, with left panels showing centerplane ($z=0$) and right panels showing axial ($y-z$) planes at $x/D=2$. Isocontours at $\bar{c}=0.2, 0.4, 0.6, 0.8$ are also shown. (a) contains LES results, (b) shows results from baseline GDH model, and (c) presents results from the TBNN-s.}
\label{fig-13-c_r1}
\end{figure}

Finally, we show a posteriori results which test the ability of the model integrated into a computational fluid dynamics code to predict the mean concentration distribution. Appendix~A includes a discussion on the relative importance of different components of the turbulent scalar flux on the a posteriori results discussed here. We compare results from two different models. The baseline is a simple, isotropic GDH where the turbulent diffusivity is prescribed with a fixed turbulent Schmidt number, $Sc_t=0.85$ \citep{kays1994turbulentPrt}. The second is the TBNN-s model which is proposed in the current paper and trained on the $r=2$ dataset. Both modelled results use the mean velocity from the LES and depend on a baseline turbulence model for the momentum equations, which is the two-equation realizable $k-\epsilon$ model in the present work.

\begin{figure}
  \begin{center}
  \includegraphics[width = 130mm]{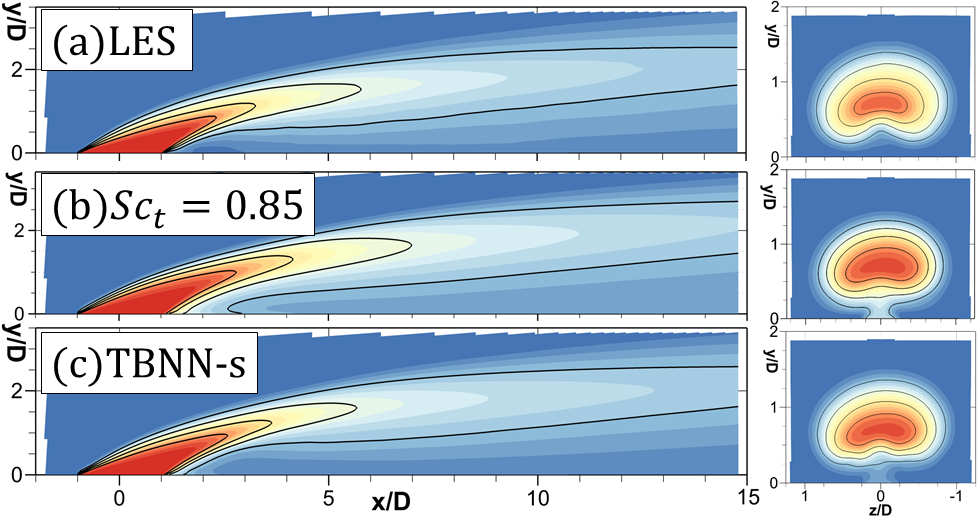}
  \end{center}
\caption{Mean scalar concentration contours in the $r=2$ case, with same legend as figure~\ref{fig-13-c_r1}.}
\label{fig-14-c_r2}
\end{figure}

Figures~\ref{fig-13-c_r1} and \ref{fig-14-c_r2} contain 2D contours of the mean scalar concentration $\bar{c}$ in the $r=1$ and $r=2$ cases respectively. In figure~\ref{fig-13-c_r1} the LES results indicate that the $r=1$ jet detaches from the wall after injection and reattaches at about $x/D=4$. The mean profile is distorted by the presence of the counter-rotating vortex pair as the axial ($y-z$) plane shows. The isotropic model, shown in figure~\ref{fig-13-c_r1}(b), broadly overestimates mixing which transports scalar towards the wall immediately downstream of injection and speeds up the decay of the jet. In the axial view, the extent of the jet is larger in the spanwise direction due to the increased mixing. The TBNN-s model, as shown in figure~\ref{fig-13-c_r1}(c) converges to a smooth solution, which is already an achievement for a tensorial and data driven model. This solution is also significantly more accurate than the baseline model of figure~\ref{fig-13-c_r1}(b): the streamwise decay of the concentration profile is predicted much more accurately and the extent of the jet's shear layer is more consistent with the LES data. As hinted by a priori results, the near-wall behavior of the TBNN-s is still lacking, even though it represents an improvement over the simple GDH. This warrants future work to adapt the TBNN-s model to regions where non-local effects dominate and lead to counter gradient transport.

\begin{figure}
  \begin{center}
    \subfloat[$r=1$]{{\includegraphics[width=65mm]{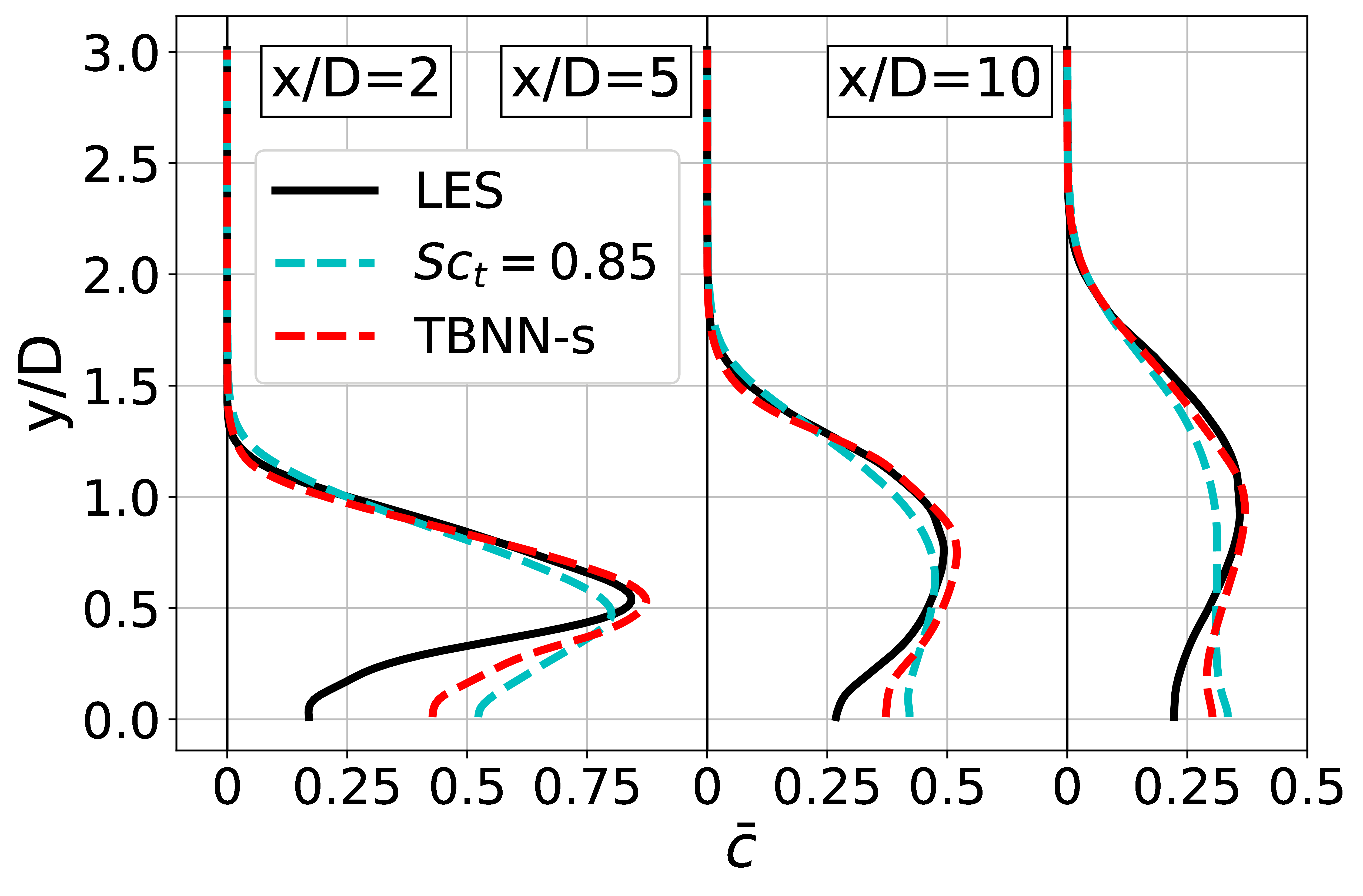} }}%
    \subfloat[$r=2$]{{\includegraphics[width=65mm]{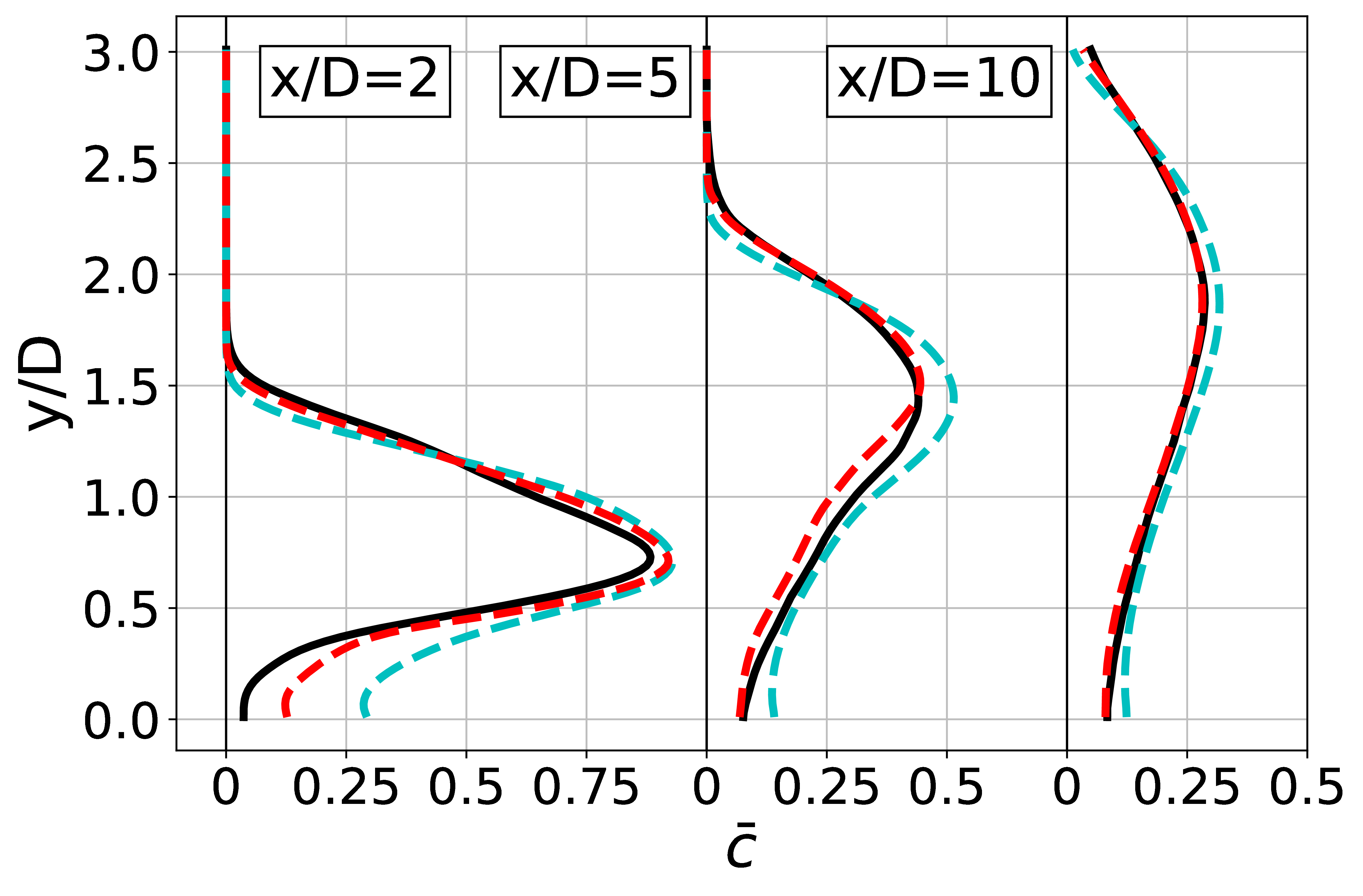} }}%
  \end{center}
\caption{Vertical profiles of $\bar{c}$ at the center spanwise position ($z=0$) calculated in three distinct streamwise locations.}
\label{fig-15-cy}
\end{figure}

Figure~\ref{fig-14-c_r2} shows the same results for the $r=2$ dataset. Here, since the jet is further from the wall, non-equilibrium effects are less significant so both models perform relatively better. The standard GDH model seems to slightly underpredict the streamwise decay of the jet, opposite from what was observed in $r=1$. The TBNN-s model in figure~\ref{fig-14-c_r2}(c) manages to correct for that. This is encouraging because by employing a tensorial diffusivity, the data driven model manages to correct the baseline model in both directions (by broadly decreasing the decay rate for $r=1$ and increasing the decay rate in $r=2$). This suggests that generalizable physics are indeed learned by the model during training and carry over to a different velocity ratio. As was observed in $r=1$, the near-wall transport, particularly right after injection, is not captured accurately. Physically, as was discussed in Section 3, this is due to the prevalence of non-equilibrium effects in this area, which cannot be modelled with a local diffusivity-based model (as evidenced by \textit{Type 2} counter gradient transport).

\begin{figure}
  \begin{center}
    \subfloat[$r=1$]{{\includegraphics[width=65mm]{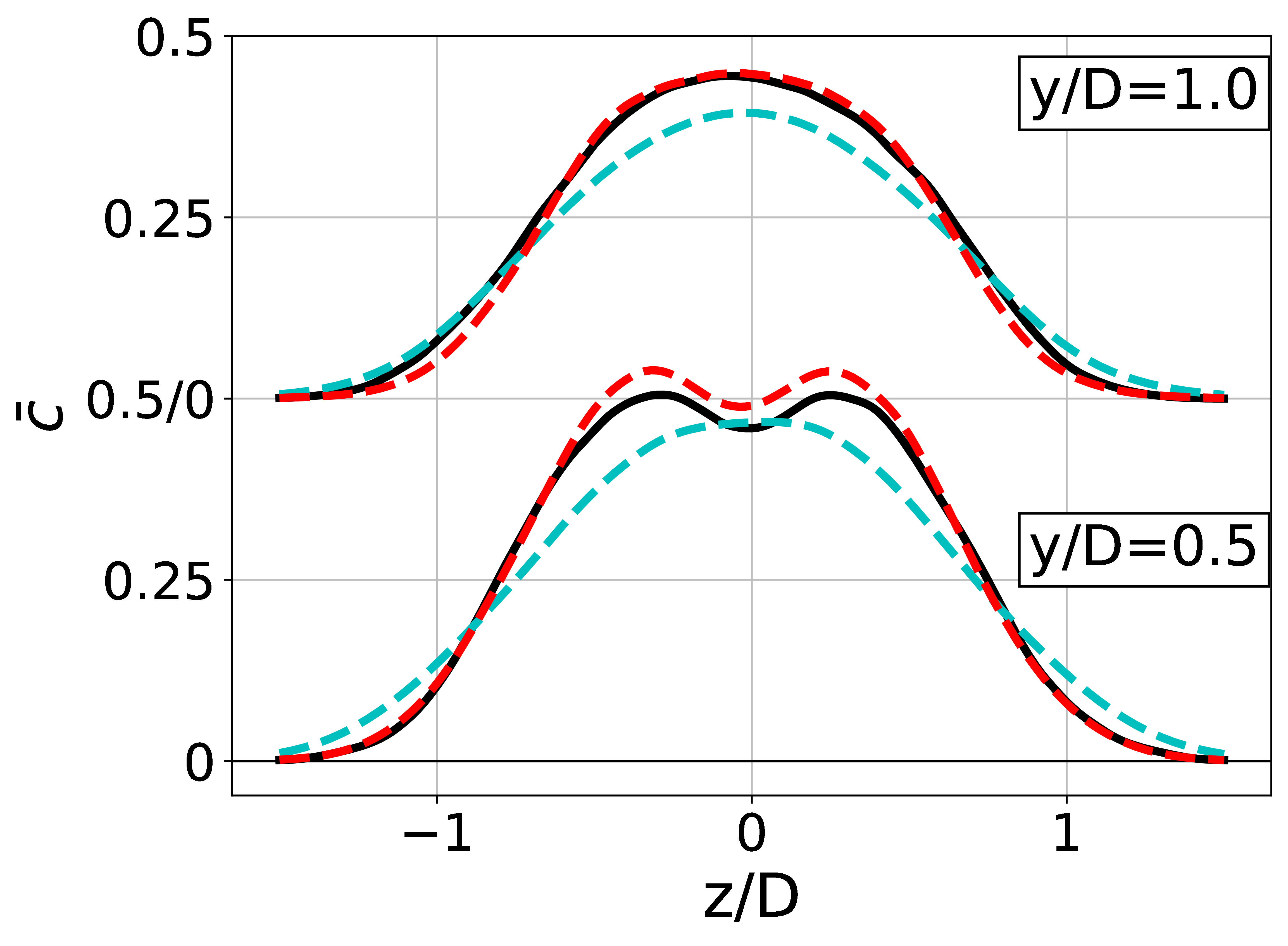} }}%
    \subfloat[$r=2$]{{\includegraphics[width=65mm]{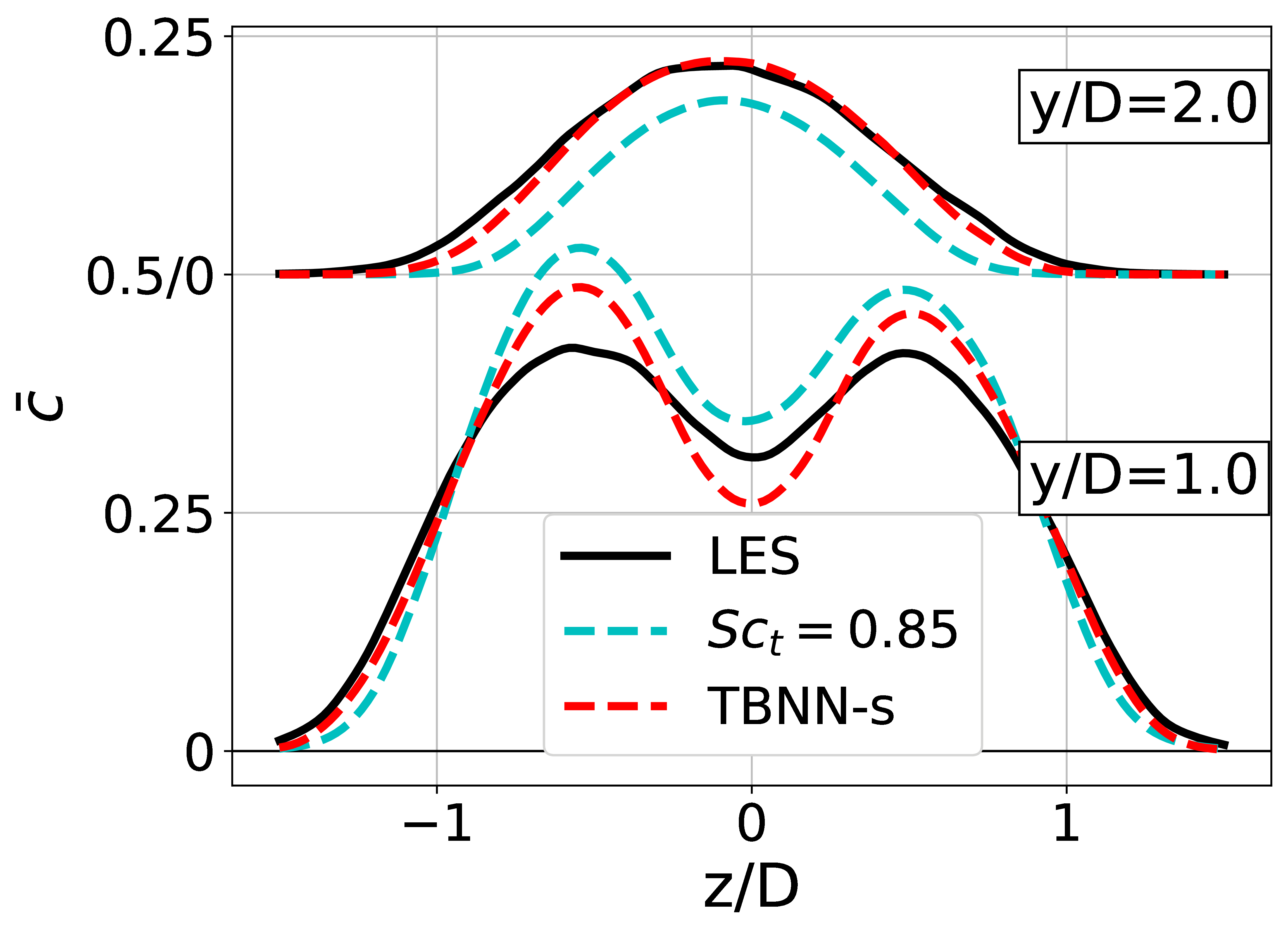} }}%
  \end{center}
\caption{Horizontal profiles of $\bar{c}$ at $x/D=5$ plotted for two distinct vertical stations.}
\label{fig-16-cz}
\end{figure}

For a more quantitative comparison between the same results, figure~\ref{fig-15-cy} shows vertical profiles of mean scalar concentration at three different streamwise locations. Despite its shortcomings, it is worth noting that the GDH with $Sc_t=0.85$ does a reasonably good job at capturing broad qualitative features of the scalar field given its simplicity. The TBNN-s model, however, presents clear improvements: it does a good job at predicting the streamwise decay of the peak jet concentration and the vertical concentration gradient in the windward shear layer. The highest discrepancy between the TBNN-s model and LES data is exactly where \textit{Type 2} gradient transport was found. Figure~\ref{fig-16-cz} shows spanwise profiles of mean scalar concentration at $x/D=5$ and further reinforces the points made in this section, especially the accuracy with which the data driven tensorial diffusivity can predict the jet shear layer profiles.

\subsubsection{Basis coefficients results}

Lastly, we would like to gain some further understanding of the TBNN-s model. Deep neural networks are notoriously powerful but hard to interpret \citep{goodfellow2016deep} and physical understanding of machine-learned turbulence model remains an important open question \citep{duraisamy2019turbulence}, so we do not intend to exhaust that question here. One method to interpret the model takes advantage of the TBNN-s structure which writes the turbulent diffusivity matrix as a weighted sum of simpler tensor basis elements. By analyzing the relative contributions of such simpler elements, one can gain useful insight into the model. Thus, in the current subsection, we look at the tensor basis weights $g^{(i)}$ prescribed by the TBNN-s in these flows. Recall that the structure in figure~\ref{fig-10-tbnns} outputs a dimensionless turbulent diffusivity matrix $\mathsfbi{D^*}$ which has O(1) entries, and the diffusivity basis elements $\mathsfbi{T^{(n)}}$ (see eq.~\ref{eq-tensorbasis_2}) are also non-dimensionalized such that in practice they have O(1) entries. Thus, the predicted values of tensor basis weights $g^{(i)}$ should be dimensionless, O(1) quantities.

\begin{figure}
  \begin{center}
  \includegraphics[width = 130mm]{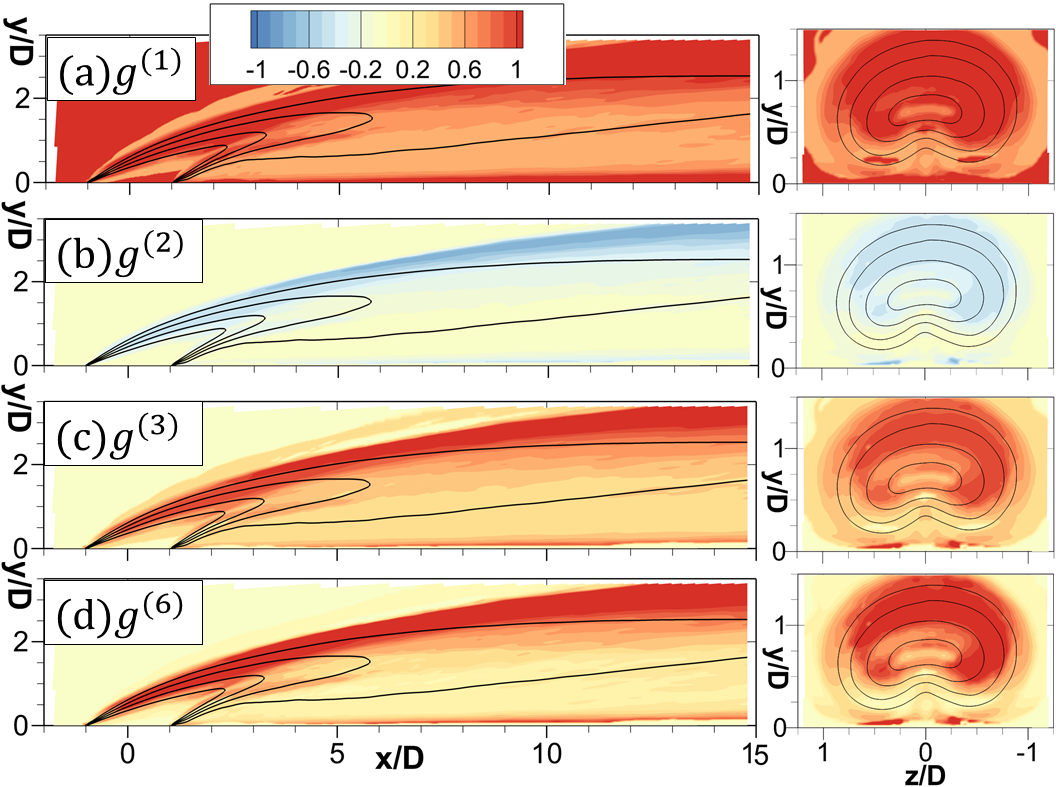}
  \end{center}
\caption{Color contours of the basis coefficients $g^{(i)}$ predicted by the TBNN-s model in the $r=2$ jet. Left shows centerplane ($z=0$) and right shows axial planes at $x/D=2$; black lines indicate isocontours of $\bar{c}=0.2, 0.4, 0.6, 0.8$.}
\label{fig-17-g_BR2}
\end{figure}

Figure~\ref{fig-17-g_BR2} contains contours of four different basis coefficients ($g^{(1)}$, $g^{(2)}$, $g^{(3)}$, and $g^{(6)}$) in the $r=2$ jet in crossflow. Since the TBNN-s was trained on the $r=2$ data, this is the machine-learned model's attempt at reproducing its training data within the tensor basis decomposition. The coefficients $g^{(4)}$ and $g^{(5)}$ are not shown because the TBNN-s assigns very low values for both of these compared to the other weights. This suggests that the quadratic and symmetric basis elements, $\mathsfbi{T}^{(4)} = \mathsfbi{S}^2$ and $\mathsfbi{T}^{(5)} = \mathsfbi{R}^2$, are not employed by the network to match the training data. All the coefficients shown have well-defined signs as learned by the deep neural network: $g^{(1)}$, $g^{(3)}$, and $g^{(6)}$ are positive, while $g^{(2)}$ is negative. Also note that while $g^{(1)}$ and $g^{(2)}$ multiply symmetric tensors, $g^{(3)}$ and $g^{(6)}$ multiply anti-symmetric basis elements. This suggests that the strictly symmetric diffusivity matrix hypothesized by some previous authors \citep[e.g.][]{daly1970transport} is insufficient to capture the data in this case, even in locations where the turbulent scalar flux is mostly locally determined such as in the windward shear layer.

\begin{figure}
  \begin{center}
  \includegraphics[width = 130mm]{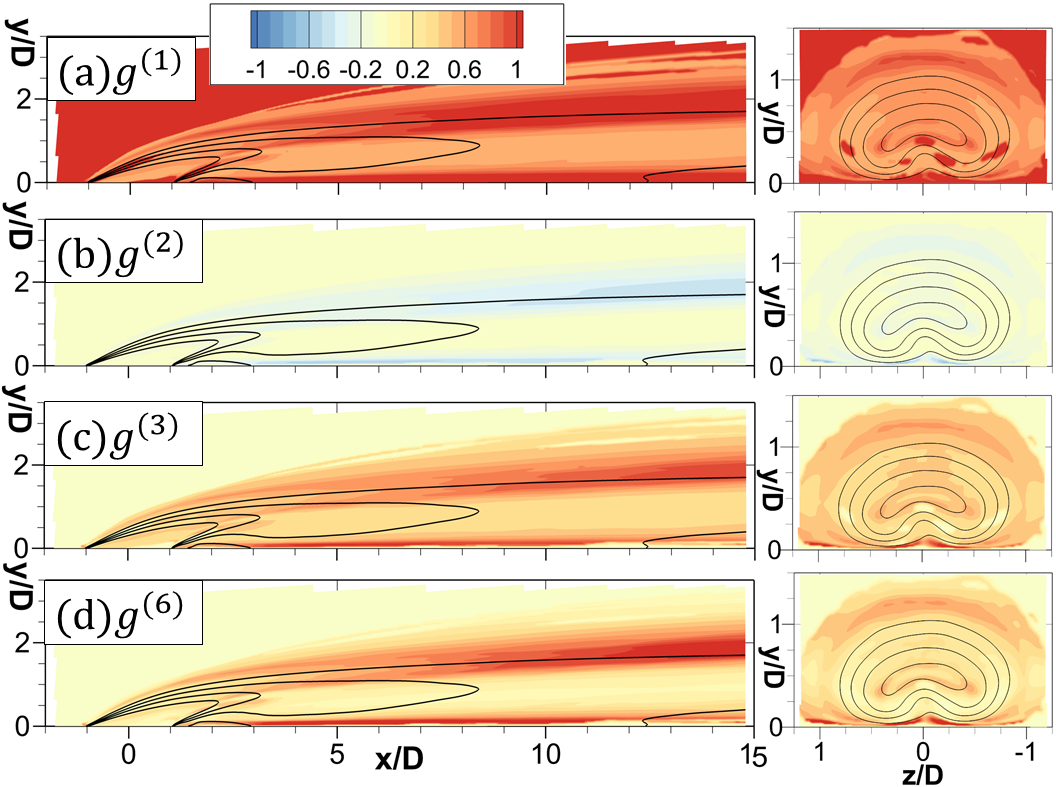}
  \end{center}
\caption{Color contours of the basis coefficients $g^{(i)}$ predicted by the TBNN-s model in the $r=1$ jet. Left shows centerplane ($z=0$) and right shows axial planes at $x/D=2$; black lines indicate isocontours of $\bar{c}=0.2, 0.4, 0.6, 0.8$.}
\label{fig-18-g_BR1}
\end{figure}

Other features can be observed in figure~\ref{fig-17-g_BR2}. For example, the basis coefficients seem to attain generally higher magnitudes in locations of highest shear, such as right above the wall and in the windward shear layer. The coefficient $g^{(1)}$ determines the isotropic contribution and is the only coefficient allowed in the simple GDH of eq.~\ref{eq-gdh}; the TBNN-s prescribes values roughly between 0.5 and 1.5 to that coefficient, with lower magnitudes in locations of \textit{Type 2} transport (compare figure~\ref{fig-9-theta_r2}(c) and figure~\ref{fig-17-g_BR2}(a)) which would be the expected behavior if the turbulent transport in those locations is to be captured. On the other hand, $g^{(1)}$ is highest in the windward shear layer, indicating that the standard GDH underestimates the isotropic mixing in those locations.

Figure~\ref{fig-18-g_BR1} shows equivalent results for the $r=1$ jet in crossflow, where the model is tested outside of its training data. The overall patterns are very similar from before; the coefficients $g^{(4)}$ and $g^{(5)}$ are significantly smaller and thus not shown. The similarity in the coefficients between the calibration data and the test data indicates that the modelling framework, in particular the tensor basis expansion, is indeed sound. One difference between the $r=1$ and $r=2$ results is that the former seem to have slightly lower magnitudes, indicating that differences in the modelled eddy viscosity $\nu_t$ are not enough to capture the increased turbulent mixing from $r=1$ to $r=2$ jet.

This opportunity for interpretation naturally arises from the tensor basis expansion described in section~4.1. The patterns indicated here can be used to further improve and simplify the models. For example, future work could remove the $\mathsfbi{T}^{(4)}$ and $\mathsfbi{T}^{(5)}$ bases, or leverage the overall structure of figures~\ref{fig-17-g_BR2}-\ref{fig-18-g_BR1} to prescribe uniform ratios $g^{(2)}/g^{(1)}$, $g^{(3)}/g^{(1)}$, and $g^{(6)}/g^{(1)}$. We also believe that interpretation work should proceed in the direction of understanding the functional form $g^{(n)}(\lambda_i)$ so that one can answer the question of why the deep neural network is predicting its results. The approaches exemplified in \citet{ling2017building} and \citet{milani2019physical} could provide a useful starting point for this endeavor. As mentioned before, this is a major open question in the field of deep learning particularly for physical applications.

\section{Concluding remarks}

Two highly resolved large eddy simulations of inclined jets in crossflow are used to analyze turbulent scalar transport. In section 3, we discuss physical origins and modelling implications of counter gradient turbulent transport. Counter gradient turbulent transport is classified into two types. \textit{Type 1} arises in the streamwise component of the turbulent scalar flux in the windward shear layer, and is caused by cross-gradient effects, whereby gradients in the wall-normal and spanwise direction control the scalar flux. This phenomenon is mostly local and is consistent with the reports of previous authors such as \citet{muppidi2008direct}, \citet{bodart2013highfidelity}, and \citet{schreivogel2016simultaneous}. \textit{Type 2} transport, on the other hand, is governed chiefly by non-local contributions to the scalar flux budget. It is observed close to the wall, after injection, and in the wall-normal component of the turbulent transport, consistent with the observations of \citet{salewski2008mixing} and \citet{milani2018magnetic}. Despite multiple reports of counter gradient transport in these flows, authors have not carefully examined its causes. For example, \citet{bodart2013highfidelity} reported observations consistent with \textit{Type 1} transport in their windward shear layer, but erroneously alluded to non-local effects as possible causes (which actually cause \textit{Type 2} transport as shown presently). These discussions show that \textit{Type 1} and \textit{Type 2} phenomena would require models of different complexities. The former could be modelled with a local formulation, but would require a tensorial diffusivity to capture cross-gradient effects. The latter cannot be modelled adequately with any diffusivity-based model since it is inherently non-local; it might require, for example, solving separate transport equations for the scalar flux components.

Given the implementation difficulties associated with non-local models \citep{combest2011gradient}, we choose to use a matrix turbulent diffusivity with a machine learning approach. We adapt the tensor basis neural network (TBNN) of \citet{julia_deepnn} to model the scalar flux, and denote this approach TBNN-s (tensor basis neural network for scalar flux modelling). Our deep learning model obeys rotational invariance through a tensor basis expansion, and is able to predict a tensorial diffusivity $\mathsfbi{D}$ that can be used in the Reynolds averaged scalar transport equations even though the training data do not contain this matrix explicitly. As expected, the turbulent diffusivity matrix improves the alignment of the turbulent scalar flux vector, and also increases the accuracy of the mean scalar concentration field. This happens not just in the training set, but also in the test flow. The TBNN-s model proposed herein can be applied in any other problem where scalar transport is of interest given adequate training data. We expect that this modelling approach could be particularly useful to practitioners who are interested in one specific class of turbulent flow and have a few high fidelity datasets to train their models. Note that unlike many other machine learning works in the recent fluid mechanics literature, this model can be directly applied to arbitrarily complex 3D turbulent flows and computational meshes. The results in this paper show that our framework is a good candidate to be employed in turbulent shear flows, where cross-gradient transport effects are important. As discussed in section~4.3.3, future workers might focus on further developing the model by simplifying it; or on further interpreting deep neural networks used in physical applications, to understand how the input invariants are mapped to output turbulence quantities of interest.

The authors would like to acknowledge Ali Mani for useful discussions and suggestions that strengthened this paper. Financial support from Honeywell Aerospace is gratefully acknowledged. Declaration of Interests. The authors report no conflict of interest.

\appendix

\section{Relative magnitude of the turbulent scalar flux}

One noteworthy aspect of modelling the turbulent scalar flux is that its contribution to the overall transport of the passive scalar varies with location and direction within the flow. In general, the passive scalar equation balances the mean advection and the turbulent scalar flux (assuming that molecular diffusion is negligible in such flows) as show in

\begin{equation}
\label{eq-raad_appendix}
\frac{\partial}{\partial x_i} (\bar{u}_i \bar{c}) = \frac{\partial}{\partial x_i} \left( - \overline{u_i'c'} \right).
\end{equation}

\begin{figure}
  \begin{center}
  \includegraphics[width = 130mm]{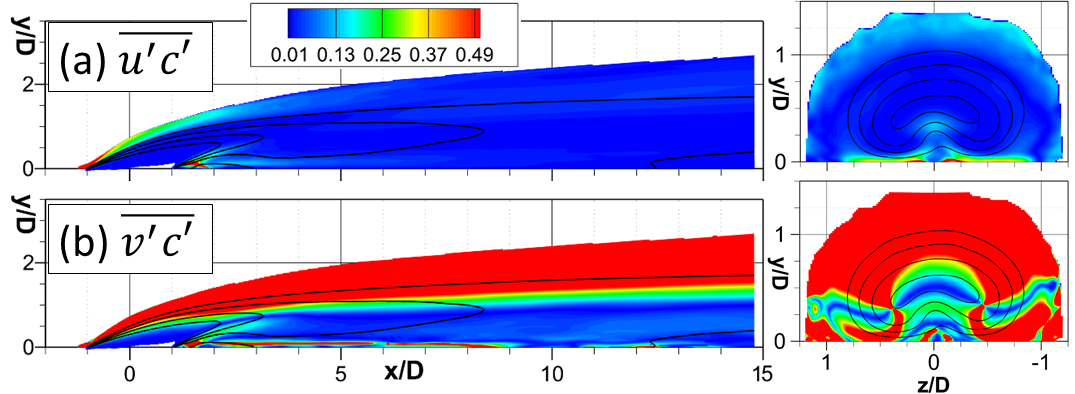}
  \end{center}
\caption{Magnitude of turbulent scalar flux components relative to the equivalent mean advection component in the $r=1$ jet. Black lines showing isocontours at $\bar{c}=0.2, 0.4, 0.6, 0.8$ are also shown. (a) contains the streamwise component, $| \overline{u'c'} / (\bar{u} \bar{c}) |$ and (b) shows the wall-normal component, $| \overline{v'c'} / (\bar{v} \bar{c}) |$.}
\label{fig-19-relmag}
\end{figure}

In shear flows such as the jet in crossflow, length scales and mean velocities in the streamwise ($x$) direction are much larger than in the transverse directions ($y$ and $z$), so streamwise mean advection is usually thought as balancing turbulent mixing in the transverse direction. This means that the computation of the mean scalar field through the RANS eq.~\ref{eq-raad_appendix} is more sensitive to errors in the transverse components of the turbulent scalar flux ($\overline{v'c'}$ and $\overline{w'c'}$) than it is to errors in the streamwise component $\overline{u'c'}$. For a visualization, figure~\ref{fig-19-relmag} shows the magnitude of the turbulent scalar flux components relative to the equivalent mean advection component from the $r=1$ jet LES. As the conceptual discussion suggests, $\overline{v'c'}$ is relatively more important than $\overline{u'c'}$; its magnitude is particularly high in the shear layer all around the jet and in some places close to the wall and right after injection.

\begin{figure}
  \begin{center}
    \subfloat[]{{\includegraphics[width=65mm]{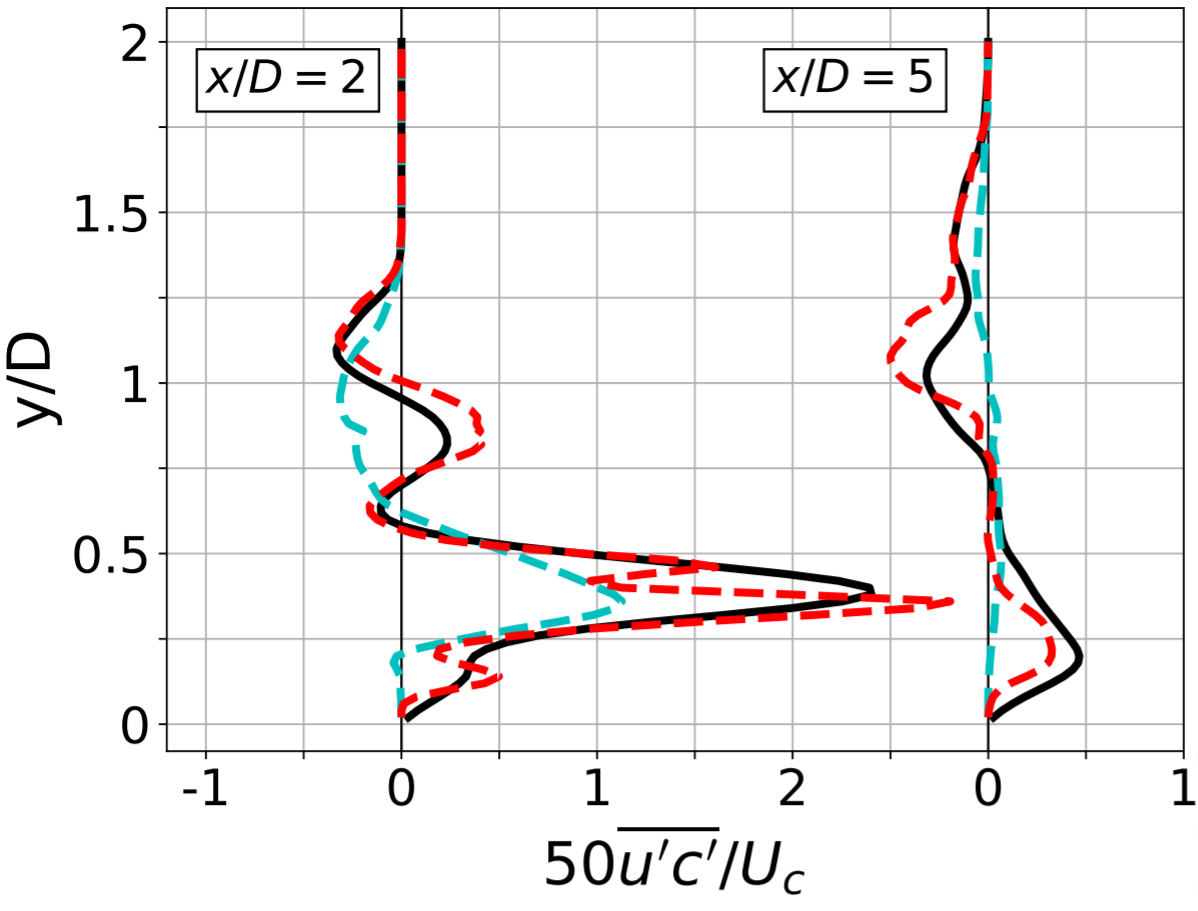} }}%
    \subfloat[]{{\includegraphics[width=65mm]{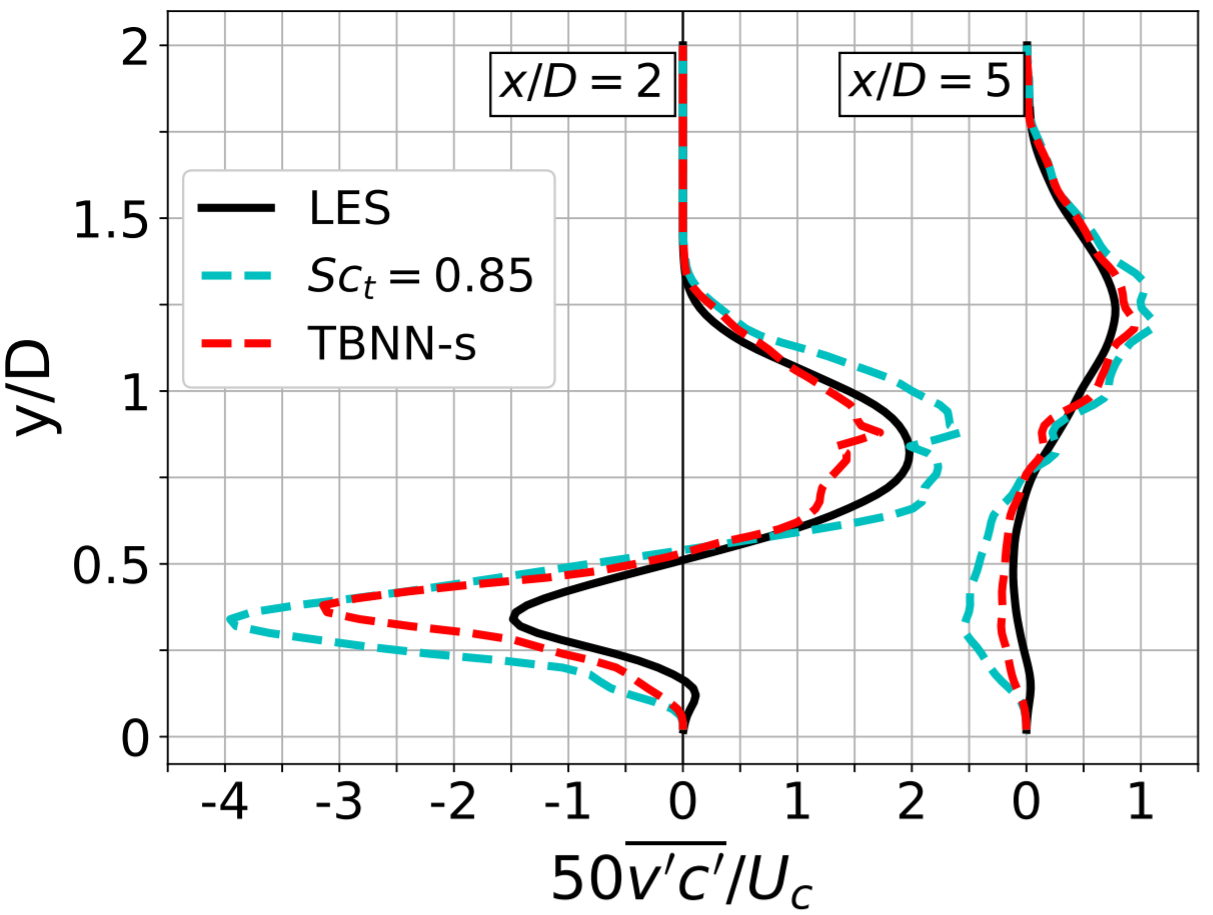} }}%
  \end{center}
\caption{Vertical profiles of the turbulent scalar flux in the $r=1$ jet at the centerplane ($z/D=0$) and two different streamwise locations. Lines compare LES results to the baseline model (GDH with $Sc_t=0.85$) and to the TBNN-s model trained on the $r=2$ jet (as discussed in section~4.3). (a) shows streamwise component $\overline{u'c'}$ of the flux, and (b) shows the wall-normal component $\overline{v'c'}$.}
\label{fig-20-uccomparison}
\end{figure}

This suggests that accurately capturing \textit{Type 1} transport does not impact significantly the predicted mean scalar field in this particular flow; on the other hand, capturing \textit{Type 2} is much more important. The goal of discussing \textit{Type 1} transport in section~3.1 is to explain the exact observations of previous authors regarding counter gradient transport \citep{muppidi2008direct, bodart2013highfidelity, schreivogel2016simultaneous} and show a specific physical mechanism that causes qualitatively incorrect predictions in the turbulent scalar flux in jets in crossflow (namely, cross-gradient transport). This physical analysis leads us to investigate machine-learned models that incorporate cross-gradient transport effects through a tensorial diffusivity, which is the focus of section~4. This improved model is not only able to capture specifically \textit{Type 1} transport (as shown in figure~\ref{fig-20-uccomparison}(a)), but it also improves the prediction of the more important components such as $\overline{v'c'}$ as shown in figure~\ref{fig-20-uccomparison}(b). This explains its significant improvement over the baseline model.

\section{Stability of a tensorial diffusivity}

In this appendix, we discuss the requirement for stability of a matrix turbulent diffusivity used to model the scalar flux in the Reynolds averaged advection diffusion equation. If the diffusivity were a scalar, it would have to be non-negative; for a matrix diffusivity, there is an analogous requirement: that the 3x3 matrix be positive semi-definite. To see this, we multiply both sides of eq.~\ref{eq-raad} by $\bar{c}$ to derive a transport equation for the squared concentration $\phi = \frac{1}{2} \bar{c}^2$.

\begin{equation}
\label{eq-raad_c}
\bar{c} \times \left[ \frac{\partial}{\partial x_i} (\bar{u}_i \bar{c}) = \frac{\partial}{\partial x_i} \left( \frac{\nu}{Sc} \frac{\partial \bar{c}}{\partial x_i} \right) + \frac{\partial}{\partial x_i} \left( D_{ij} \frac{\partial \bar{c}}{\partial x_j} \right) \right]
\end{equation}

The first term in eq.~\ref{eq-raad_c} can easily be manipulated since we assume the mean velocity field is divergence free, and simply becomes an advection of $\phi = \frac{1}{2} \bar{c}^2$. The two terms on the left hand side require a slightly more sophisticated application of the chain rule of differentiation, namely

\begin{equation}
\label{eq-chain_rule_1}
\bar{c} \frac{\partial}{\partial x_i} \left( \frac{\nu}{Sc} \frac{\partial \bar{c}}{\partial x_i} \right) = \frac{\partial}{\partial x_i} \left( \frac{\nu}{Sc} \frac{\partial \left( \frac{1}{2} \bar{c}^2 \right)}{\partial x_i} \right) - \frac{\nu}{Sc} \left(\frac{\partial \bar{c}}{\partial x_i} \frac{\partial \bar{c}}{\partial x_i} \right) \text{and}
\end{equation}

\begin{equation}
\label{eq-chain_rule_2}
\bar{c} \frac{\partial}{\partial x_i} \left( D_{ij} \frac{\partial \bar{c}}{\partial x_j} \right) = \frac{\partial}{\partial x_i} \left( D_{ij} \frac{\partial \left( \frac{1}{2} \bar{c}^2 \right)}{\partial x_j} \right) - D_{ij} \left(\frac{\partial \bar{c}}{\partial x_i} \frac{\partial \bar{c}}{\partial x_j} \right).
\end{equation}

Now, using eqs.~\ref{eq-chain_rule_1}-\ref{eq-chain_rule_2} we rewrite eq.~\ref{eq-raad_c} as

\begin{equation}
\label{eq-raad_phi}
\frac{\partial}{\partial x_i} (\bar{u}_i \phi) = \frac{\partial}{\partial x_i} \left( \frac{\nu}{Sc} \frac{\partial \phi}{\partial x_i} \right) + \frac{\partial}{\partial x_i} \left( D_{ij} \frac{\partial \phi}{\partial x_j} \right) - \frac{\nu}{Sc} \left(\frac{\partial \bar{c}}{\partial x_i} \frac{\partial \bar{c}}{\partial x_i} \right) - D_{ij} \left(\frac{\partial \bar{c}}{\partial x_i} \frac{\partial \bar{c}}{\partial x_j} \right).
\end{equation}

Note that the first three terms of eq.~\ref{eq-raad_phi} are the regular mean advection and diffusion terms of quantity $\phi$. The last two act as source terms. For the diffusion of $c$ to be stable, it must act to dissipate the quantity $\phi$ (the same way that momentum diffusion acts to dissipate kinetic energy), thus the source terms must be non-positive. For the molecular diffusion, this is true if the diffusion coefficient $\frac{\nu}{Sc}$ is non-negative. The last term of eq.~\ref{eq-raad_phi} is a quadratic form of matrix $\mathsfbi{D}$ and will be non-positive if $\nabla \bar{c}^T \mathsfbi{D} \nabla \bar{c} \geq 0$. For this to hold for any mean concentration field, we require that $\mathsfbi{D}$ be positive semi-definite (which by definition means that the quadratic form of $\mathsfbi{D}$ with any vector is non-negative).

The remaining question is, given an arbitrary matrix diffusivity $\mathsfbi{D}$, how to test whether it is positive semi-definite? For a symmetric 3x3 matrix, the simplest test lies in its eigenvalues: a symmetric matrix is positive semi-definite if and only if all its eingenvalues are non-negative. For a general matrix, with complex eigenvalues, one extra step is needed. First, we split $\mathsfbi{D}$ in symmetric and anti-symmetric components, $\mathsfbi{A}$ and $\mathsfbi{W}$ respectively:

\begin{equation}
\label{eq-sym_asym}
\mathsfbi{D} = \overbrace{\frac{1}{2} (\mathsfbi{D} + \mathsfbi{D}^T)}^\mathsfbi{A} + \overbrace{\frac{1}{2} (\mathsfbi{D} - \mathsfbi{D}^T)}^\mathsfbi{W}.
\end{equation}

\noindent Then, we recognize that the quadratic form of an anti-symmetric matrix with any real vector is zero. Thus, the quadratic form of $\mathsfbi{D}$ with an arbitrary vector $\boldsymbol{v}$ is identical to that of its symmetric part $\mathsfbi{A} = \frac{1}{2} (\mathsfbi{D} + \mathsfbi{D}^T)$ with $\boldsymbol{v}$, as shown in eq.~\ref{eq-quadform}. From that, we see that $\mathsfbi{D}$ is positive semi-definite if and only if its symmetric part $\mathsfbi{A}$ is positive semi-definite.

\begin{equation}
\label{eq-quadform}
\boldsymbol{v}^T \mathsfbi{D} \boldsymbol{v} = \boldsymbol{v}^T \mathsfbi{A} \boldsymbol{v} + \cancelto{0}{\boldsymbol{v}^T \mathsfbi{W} \boldsymbol{v}}
\end{equation}

In summary, to check whether an arbitrary turbulent diffusivity matrix $\mathsfbi{D}$ is positive semi-definite, we look at its symmetric part: $\mathsfbi{D}$ will be positive semi-definite if and only if $\frac{1}{2} (\mathsfbi{D} + \mathsfbi{D}^T)$ contains only non-negative eigenvalues. This criterium is used to post process the output of the TBNN-s, which leads to a stable advection diffusion equation with a matrix turbulent diffusivity. Note that having only non-negative diagonal entries is a necessary but not sufficient condition for $\mathsfbi{D}$ to be positive semi-definite. On the other hand, a positive semi-definite matrix can have positive, zero, or negative off-diagonal entries.

\bibliographystyle{jfm}

\bibliography{milani-bib}

\end{document}